\documentclass[journal]{IEEEtran}

\usepackage{amssymb, graphicx, cite, amsmath, psfrag, minibox, mathrsfs, bigints, stfloats, color, upgreek, gensymb, epstopdf, textcomp, amsthm, amsfonts, euscript, calligra, xcolor, pict2e, tikz}

\usepackage{hyperref}

\usepackage[font=scriptsize]{subcaption}
\usepackage[font=scriptsize]{caption}

\usepackage{mathtools, amsfonts, multirow, array}
\usepackage{verbatim, mdwtab, relsize, cleveref, mdwmath}
\usepackage{algorithmicx}
\usepackage{algpseudocode}
\usepackage[ruled]{algorithm2e}
\usepackage{collectbox}
\usepackage{comment}

\usepackage{setspace}
\usepackage{float}

\usepackage{tikz}
\usetikzlibrary{shapes,arrows}
\usepackage{textcomp}

\definecolor{col1}{rgb}{0.2,0.3,0.8}
\definecolor{col2}{rgb}{0.5,0.1,0.1}
\definecolor{col3}{rgb}{0.302, 0.745, 0.9}

\makeatletter
\newcommand{\leqnomode}{\tagsleft@true}
\newcommand{\reqnomode}{\tagsleft@false}
\makeatother

\def\Ibar{\overline{I}_{\rm{av}}}
\def\SUtx{SU$_{\rm{tx}}$}
\def\SUrx{SU$_{\rm{rx}}$}
\def\PUtx{PU$_{\rm{tx}}$}
\def\PUrx{PU$_{\rm{rx}}$}

\def\pu{p_{\rm u}}
\def\Nd{N_{\rm d}}
\def\Td{\tau_{\rm d}}

\def\Dd{D_{\rm d}}
\def\Dt{D_{\rm t}}
\def\Pp{P_{ p}}
\def\eu{e_{\rm u}}
\def\Nu{N_{\rm u}}

\def\fs{f_{\rm s}}
\def\Tf{T_{\rm f}}
\def\Tsen{\tau_{\rm s}}
\def\Nsen{N_{\rm s}}
\def\Ttra{\tau_{\rm{t}}}
\def\Ptra{P_{\rm t}}
\def\Ntra{N_{\rm t}}
\def\alphatra{\alpha_{\rm t}}
\def\RLB{R_{\rm{LB}}}

\def\gamn{ \gamma_n }
\def\gamhatn{ \widehat{\gamma}_n }
\def\ghatn{ \widehat{g}_n }
\def\gamtiln{ \widetilde{\gamma}_n }

\def\Sigmap{ \sigma_{ p} }
\def\Sigmawn{ \sigma_{{ w}_n} }
\def\Sigmavn{ \sigma_{{ v}_n} }
\def\piohn{ \widehat{\pi}_{0,n} }
\def\pilhn{\widehat{\pi}_{1,n} }
\def\omegaon{\omega_{0,n} }
\def\omegaln{\omega_{1,n} }
\def\betaon{\beta_{0,n} }
\def\betaln{\beta_{1,n} }

\def\Ho{ {\cal H}_0 }
\def\Hl{ {\cal H}_1 }
\def\Hhaton{ \widehat{\cal H}_{0,n} }
\def\Hhatln{ \widehat{\cal H}_{1,n} }
\def\Heps{ {\cal H}_\varepsilon }

\ifCLASSINFOpdf
\else
\fi

\begin{document}
\title{ Steady-State Rate-Optimal Power Adaptation in Energy Harvesting Opportunistic Cognitive Radios with Spectrum Sensing and Channel Estimation Errors
}

\author{Hassan Yazdani, Azadeh~Vosoughi,~\IEEEmembership{Senior Member,~IEEE} 
    	\thanks{This research was supported by NSF under grant CISE-CNS-2006683.} 
}

\markboth{}%
{Shell \MakeLowercase{\textit{et al.}}: Bare Demo of IEEEtran.cls for Journals}
%

\newcounter{MYeqcounter}
\multlinegap 0.0pt                     

\maketitle

 
\begin{abstract}
We consider an opportunistic  cognitive radio network,  consisting of $\Nu$  secondary users (SUs) and an access point (AP), that can access a spectrum band licensed to a primary user. Each SU is capable of harvesting energy, and is equipped with a finite size  battery, for energy storage. The SUs operate under a time-slotted scheme, where each time slot consists of three non-overlapping phases: spectrum sensing phase, channel probing phase, and data transmission phase.  The AP feeds back its estimates of fading coefficients of SUs--AP link to SUs. To strike a balance between the energy harvesting and the energy consumption, we propose a parametrized power control strategy that allows each SU to adapt its power, according to the feedback information and its stored energy. Modeling  the randomly arriving energy packets during a time slot as a Poisson process, we establish a lower bound on the achievable sum-rate of SUs--AP links, in the presence of both spectrum sensing and channel estimation errors. We optimize the parameters of the proposed power control strategy, such that the derived sum-rate lower bound is maximized, subject to an interference constraint. Via simulations, we corroborate our analysis and explore spectrum sensing-channel probing-data transmission trade-offs. 
\end{abstract}
%

\IEEEpeerreviewmaketitle

\section{Introduction}\label{Se1}








\subsection{Literature Review}

\par The explosive rise in demand for high data rate wireless applications has turned the spectrum into a scarce resource. Cognitive radio (CR) technology is a promising solution which alleviates spectrum scarcity problem by allowing an unlicensed secondary user (SU) to access licensed frequency  bands in a such way that its imposed interference on primary users (PUs) is limited \cite{Arslan, Hamouda_Survey, R17}. Therefore, CR systems can increase spectrum efficiency significantly. CR systems are mainly classified as  underlay CR and opportunistic CR systems. In underlay CR systems, SUs use a licensed frequency band simultaneously with PUs, conditioned that the interference power imposed on PUs (caused by SUs) remains below a pre-determined level. 
In opportunistic CR systems, SUs use a licensed frequency band as long as the frequency band is not used by PUs. While opportunistic CR systems do not require  coordination between PUs and SUs to acquire channel state information (CSI) corresponding to SU-PU link, they necessitate spectrum sensing to monitor and detect PUs' activities and protect PUs against harmful interference caused by SUs  \cite{CISS2019, J1, J2}. In these systems, the status of PUs' activities (being busy or idle) and the duration of spectrum sensing affect the system performance \cite{J1, J2, Cheung, Yin, Biswas}.
Spectrum sensing is prone to errors (characterized in terms of mis-detection and false alarm probabilities) which need to be considered in the system design. Another important factor that impacts the performance of opportunistic CR systems is the level of assumption made regarding the availability of CSI. In opportunistic CR systems, although CSI corresponding to SU-PU link is not required (which is a major advantage), still CSI corresponding to SU transmitter-- SU receiver (\SUtx--\SUrx) link is needed for properly adapting the data transmission.

\par In addition to spectral efficiency, energy efficiency is another important metric to consider when designing  communication systems \cite{EH_Survey, Nalla, Liang_Energy, EH_Probing, Power_Control, Sultan, FanConf}. Energy harvesting (EH) has been recognized as an effective approach for improving the energy efficiency. EH-powered devices can operate without the need for external power cables or periodic battery replacements \cite{Ghazaleh1, ardeshiri2019power2}. EH-enabled  CR systems have received substantial attention as a promising solution for increasing both energy efficiency and spectral efficiency \cite{EHCRN1, Park_EH, zhang2017energy}.  
EH-enabled communication systems can harvest energy from ambient energy sources (e.g., solar, wind, thermal, vibration) or radio frequency (RF) signals. {For instance, in  an ambient RF EH-enabled CR system, the energy of emitted RF signals from TV/radio broadcast towers, cellular base stations, and Wi-Fi access points (APs) is captured by \SUtx ~antenna and stored in its battery \cite{Lee2, Niyato, Finite_Markov, Spatial_Throughput, Hong_PIMRC}. A dedicated RF signal source can be utilized for energy harvesting and enabling simultaneous wireless information and power transfer (SWIPT) \cite{Dynamic_SWIPT, Cheng_SWIPT}.

\par The body of research on EH-enabled communication systems can be grouped  into two main categories, depending on the adopted energy arrival model \cite{EH_Survey, Alouini_EH}: in the first model, the energy arrival is deterministic and the transmitter has a causal or non-causal knowledge of the energy arrival at the beginning of transmission \cite{limited_feedback}. In the second  model, the energy arrival is stochastic \cite{EH_Survey}. In practice, the energy arrival of ambient energy sources, including ambient RF signal sources, is intrinsically  time-variant and often sporadic. This natural  factor degrades the performance of the battery-free EH-enabled communication systems in which a “harvest-then-transmit” strategy is adopted, i.e., users can only transmit when the energy harvested in one time slot is sufficient for data transmission \cite{Sakr}. To flatten the randomness of the energy arrival, the harvested energy is stored in a battery, to balance the energy arrival and the energy consumption \cite{EH_Survey}. In practice, the capacity of the batteries is limited, and this can result in an energy overflow.

\par Power/energy management in EH-enabled communication  systems with finite size batteries is necessary, in order  to adapt the rate of energy consumption with the rate of {energy harvesting}. If the energy management policy is overly aggressive, such that the rate of energy consumption is greater than the rate of energy harvesting, the transmitter may stop functioning, due to energy outage. On the other hand, if the energy management policy is overly conservative, the transmitter may fail to utilize the excess energy, due to energy overflow, and the data transmission would become limited in each energy allocation interval.

\par Focusing on opportunistic EH-enabled CR systems, we realize that power control strategies, aiming at optimizing the performance of SUs, should be designed such that spectrum sensing (and its corresponding errors), as well as spectrum sensing-data transmission trade-offs are incorporated in the design process. For instance, the authors in \cite{Lee2} considered a system model, where \SUtx ~can perform energy harvesting and spectrum sensing  simultaneously. Depending  on the results of spectrum sensing, \SUtx ~continues to harvest energy (when the spectrum is sensed busy) or transmits data (when the spectrum is sensed idle), and studied maximizing \SUtx--\SUrx ~channel capacity, via optimizing the threshold of the energy detector (employed for spectrum sensing). 
Aiming at a similar goal (i.e., maximizing the SU's channel capacity), the authors in \cite{Hong_PIMRC} considered a modified system model, where \SUtx ~cannot perform energy harvesting and spectrum sensing at the same time. The authors investigated the optimal mode selection policy (i.e., to choose whether to access the spectrum or to harvest energy) for CR sensor networks. Targeting the same goal as\cite{Lee2, Hong_PIMRC}, the authors in \cite{Cheung, FanConf} studied the optimal allocation of energy to be consumed for spectrum sensing versus data transmission, assuming that \SUtx ~has a finite size data buffer.
The authors in  \cite{Yin} considered a different system model, where energy harvesting, spectrum sensing, and data transmission occur in three non-overlapping time intervals within a frame. They studied maximizing the \SUtx--\SUrx ~link throughput, via optimizing the duration of spectrum sensing and  the threshold of the energy detector (for spectrum sensing), and investigated the energy harvesting-spectrum sensing-data transmission trade-offs. We note that the works in \cite{Hong_PIMRC, Cheung, FanConf, Lee2, Yin} assume that CSI of \SUtx--\SUrx ~link is perfectly known at both \SUtx ~and \SUrx.

\par In general, the power control strategies designed for opportunistic EH-enabled CR systems should depend on the level of assumption made regarding the availability of CSI corresponding to \SUtx--\SUrx ~link, and whether the adapted transmit power levels are continuous or discrete values. In practice, only partial CSI can be available at \SUtx ~and \SUrx ~due to several factors (e.g., channel estimation error and limitation of feedback channel from \SUrx ~to \SUtx). Partial CSI has deteriorating effects on the performance of communication systems (including EH-enabled CR systems),  and should not be overlooked. 
In the following, we reference several works that consider the effects of partial CSI on the performance of EH-enabled communication systems, with stochastic energy arrival model and finite size batteries. Assuming perfect CSI at the receiver (Rx) and quantized CSI  (due to limited feedback channel) at the transmitter (Tx), the authors in \cite{Data_Driven} aimed at maximizing the Tx's bit rate, via adapting discrete-valued data transmit power and modulation order, according to the quantized CSI of  Tx--Rx link and the Tx battery state. 
Assuming perfect CSI at the Rx and single-bit partial CSI at the Tx (due to severely limited feedback channel), the authors in \cite{limited_feedback} targeted at maximizing the Tx-Rx link throughput, via optimizing the {threshold of the binary channel quantizer} and discrete-valued data transmit power. 
Assuming perfect CSI at the Rx and partial CSI at the  Tx (due to channel estimation error), the authors in \cite{Alouini_EH, Zenaidi} analyzed maximizing the Tx's average throughput, in two asymptotic regimes (where the rate of energy harvesting is very small or very large), via optimizing continuous-valued data transmit power. 
We note that, none of the referenced works in \cite{Data_Driven, Alouini_EH, Zenaidi, limited_feedback} considered CR systems. Furthermore, these works assume that CSI is perfectly known at the Rx.
%
%
\subsection{Knowledge Gap and Our Contributions}

\par We consider an opportunistic EH-enabled CR network, consisting of $\Nu$ SUs and an access point (AP), that can access a wideband spectrum licensed to a primary network. Each SU is capable of harvesting energy from natural ambient energy sources, and is equipped with a finite size rechargeable battery, to store the harvested energy. Our main {\it objectives} are (i) to study how the achievable sum-rate of SUs is impacted by the {\it combined effects} of spectrum sensing error and imperfect CSI of SUs--AP links  (due to channel estimation error), and (ii) to design an energy management strategy that maximizes the achievable sum-rate of SUs, subject to a constraint on the average interference power that SUs can impose on the PU. To the best of our knowledge, our work is the first to study the impact of these combined effects on the performance of an opportunistic EH-enabled CR network. 

\par The importance of our study is evident by the works in \cite{Hassibi1, Medard, Shirazi2, Ahmadi2, Vosoughi_Capacity, Jia2}, which demonstrate the significance of considering the effect of imperfect CSI at the Rx, due to channel estimation, on the Tx achievable rate. We note that the Tx in these works is a primary transmitter (not a secondary transmitter in a CR system) and has a traditional stable power supply. One expects that spectrum sensing error, combined with random energy arrival at the Tx, exacerbates the effect of imperfect CSI on the Tx achievable rate. The challenges of our study are twofold:  first, it requires integration of energy harvesting, spectrum sensing, and channel estimation. Successful achievement of this integration entails stochastic modeling of energy arrival, energy storage, and PU's activities.  These stochastic models are utilized to establish an achievable sum-rate of SUs that takes into account both spectrum sensing error and channel estimation error. Second, one needs to properly design energy control strategies for SUs, that strike a balance between the energy harvesting and the energy consumption, and adapt transmit power according to the available CSI and the battery state. 

\par We assume that SUs operate under a time-slotted scheme, and SU$_n$ is capable of harvesting energy during the entire time slot. Each time slot consists of three sub-slots corresponding to spectrum sensing phase (during which SU$_n$ senses the spectrum), channel probing phase (during which SU$_n$ sends pilot symbols to the AP, when the spectrum is sensed idle, for estimating the fading coefficient corresponding to SU$_n$-AP link), and data transmission phase (during which SU$_n$ sends data symbols to the AP). Assuming that the AP feeds back its estimate of the fading coefficient to SU$_n$, SU$_n$ adapts its transmit power based on this information as well as the available energy in its battery.

\par Our main contributions can be summarized as follow:

\noindent 1) Our system model encompasses the stochastic energy arrival model for harvesting energy, the stochastic energy storage model for the finite size battery, the stochastic model of PU's activities, spectrum sensing error, and channel estimation error (both at SUs and the AP). We model  the randomly arriving energy packets during a time slot as a Poisson process, and the dynamics
of the battery as a finite state Markov chain. 

\noindent 2) We propose a power adaptation strategy for SU$_n$ that mimics the behavior of the rate-optimal power adaptation scheme with respect to the estimated channel power gain $\ghatn$ available at SU$_n$ and the AP, i.e., when $\ghatn$ is below a cut-off threshold $\theta_n$, the transmit  energy is zero, and when $\ghatn$  exceeds $\theta_n$, the transmit energy increases monotonically in proportion to a parameter $\Omega_n$, as $\ghatn$ increases. The parameters $\Omega_n$ and $\theta_n$ play key roles  in balancing the energy harvesting and the energy consumption.

\noindent 3) Given our system model, we establish a lower bound on the achievable sum-rate of SUs-AP links, in the presence of both spectrum sensing error and channel estimation error (both at SUs and the AP). We formulate a novel constrained optimization problem with the optimization variables $\{\Omega_n,\theta_n \} _{n=1}^ {\Nu}$, aiming at maximizing the derived sum-rate lower bound, subject to the average interference constraint (AIC) imposed on the PU and the causality constraint of the battery. We solve the formulated constrained optimization problem assuming that the battery reaches its steady-state.

\noindent 4) We derive closed form expressions for the battery outage probability and transmission outage probability and demonstrate their behaviors, in terms of the average number of harvesting energy packets and the AIC. We also study the existing trade-offs between spectrum sensing-channel probing-data transmission and how these trade-offs impact the sum-rate of our CR network.

\par Our work is different from \cite{Alouini_EH, limited_feedback, Data_Driven, Sakr}. In particular, these works view the energy management policy design as a sequential decision making problem, and hence, they adopt the Markov decision process (MDP) framework to solve the problem. In this framework, the goal is typically optimizing a specific metric over a horizon spanning several time slots. The solutions (obtained using dynamic programing) are dependent across time slots, and also depend on the initial condition (i.e., the initial state of the battery). Here, we assume that the battery operates at its steady-state, and hence, our proposed constrained optimization problem can be solved for each time slot. Furthermore, the problem can be solved {\it offline} and the optimized transmission parameters $\{\Omega_n,\theta_n\}_{n=1}^{\Nu}$ (that do {\it not} depend on the initial condition of the battery)  can become available {\it apriori} at the AP and SUs. During the data transmission phase, SU$_n$ chooses its symbol power, using its optimized transmission parameters $\Omega_n,\theta_n$, and based on its partial CSI of SU$_n$-AP link (received via the feedback channel) as well as the available energy in its battery. 
%
%
\subsection {Paper Organization}

\par The remainder of the paper is organized as follows. Section \ref{SyS_Model} explains our system model. Section \ref{EnergyDetector} describes the spectrum sensing phase and  our binary energy-based detector for detecting PU activity. Section \ref{Ch_Training} discusses the channel probing phase. Section \ref{Phase3} explains the data transmission phase and derives a lower bound on the achievable sum-rate of our CR network. Also, it formulates our proposed constrained optimization problem. Section \ref{SimResults} corroborates our analysis on the proposed optimization problem with Matlab simulations. Section \ref{Conclu} concludes the paper.
%
\section{System Model} \label{SyS_Model}
%
%
\begin{figure}[!t]
\centering 
\hspace{-2mm}
\scalebox{1}{
\begin{picture}(180,180)
{ \scriptsize 
	\put(15,0){\includegraphics[width=60mm]{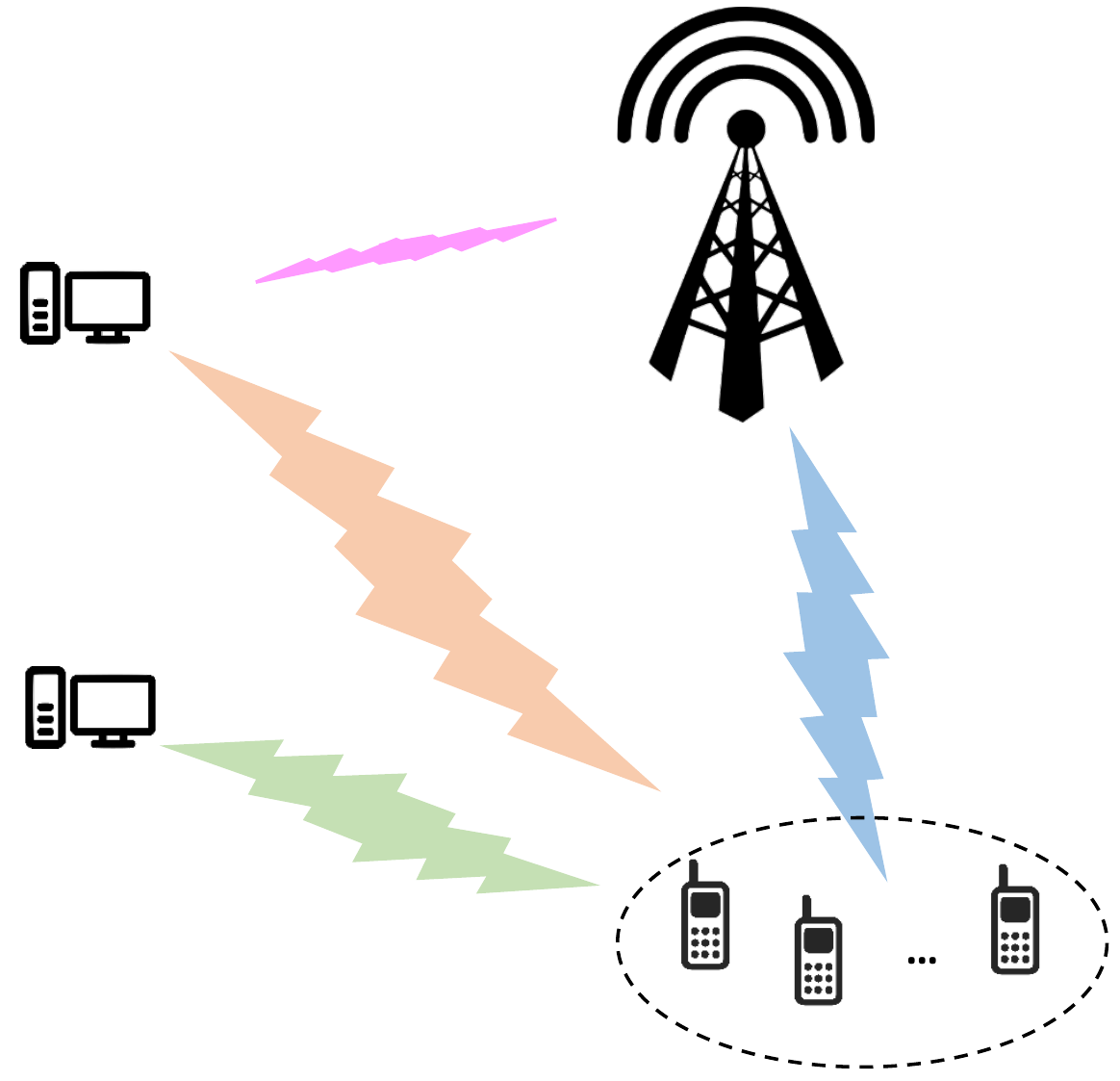}}
	\put(68,40){$\boldsymbol{z}$}
	\put(75,75){$\boldsymbol{u}$}
	\put(140,65){$\boldsymbol{h}$}
	\put(75,130){$q$}
	\put(22,67){\PUrx}
	\put(22,130){\PUtx}
	\put(145,118){AP} }
{ \tiny
	\put(119,10){SU$_1$}
	\put(136,5){SU$_2$}
	\put(165,10){SU$_{\Nu}$} }
\end{picture}
}
\caption{Schematics of the uplink CR network.} 
\label{SysModel1}
\vspace{0mm}	
\end{figure}
%
%
\par We consider an uplink opportunistic CR network that can access a wideband spectrum band licensed to a primary network, consisting of $M$ non-overlapping narrowband spectrum bands, each with a bandwidth of $W$ Hz \cite{R17}.
The primary network consists of a primary transmitter (\PUtx) and a primary receiver (\PUrx). The secondary network consists of an AP and $\Nu$ SUs (see Fig. \ref{SysModel1}). The AP can serve up to $M$ SUs simultaneously and we assume that $\Nu \leq M$. We  also assume that narrowband spectrum bands are pre-assigned to SUs and thus each SU knows which band to sense and transmit data over. The SUs are equipped with identical energy harvesting circuits to harvest energy from the ambient environment and identical finite size batteries for energy storage (see Fig. \ref{CRStructure}). We consider block fading channel model and suppose flat fading coefficients  from \PUtx ~to SU$_n$, \PUtx ~to AP,  SU$_n$ to \PUrx, and,  SU$_n$ to AP are four independent zero-mean complex Gaussian random variables, which we denote by $u_n$, $q$, $z_n$ and $h_n$  with variances $\delta_{ u_n}$, $\delta_{q}$, $\delta_{ z_n}$ and $\gamma_{n}$, respectively.  
%
%
 \begin{figure}[!t]
 \centering
 \hspace{-6mm}
 \scalebox{0.75}{
 \begin{tikzpicture}
 \draw [fill=red!20] (-2,0) circle [radius=0.5];
 \draw [fill=red!20] (2,0) circle [radius=0.5];
 \draw [fill=blue!20] (-2,-2) circle [radius=0.5];
 \draw [fill=blue!20] (2,-2) circle [radius=0.5];
 \node at (-1.95,0) {PU$_{\rm tx}$};
 \node at (2.05,0) {PU$_{\rm rx}$};
 \node at (-1.95,-2) {SU$_n$};
 \node at (2.00,-2) {AP};
 \draw [dashed, thick, ->] (-1.5,-2) -- (1.5,-2);
 \draw [dashed, thick, ->] (-1.5,0) -- (1.5,-2);
 \draw [dashed, thick, ->] (-1.5,-2) -- (1.5,0);
 \draw [dashed, thick, ->] (-2,-0.5) -- (-2,-1.5);
 \draw [thick, -] (-3.8,-2) -- (-2.5,-2);
 \draw [thick, fill=orange!60] (-4,-1.5) rectangle (-3.6,-0.5);
 \draw [thick, -] (-4,-1.3) -- (-3.6,-1.3);
 \draw [thick, -] (-4,-1.1) -- (-3.6,-1.1);
 \draw [thick, -] (-4,-0.9) -- (-3.6,-0.9);
 \draw [thick, -] (-4,-0.7) -- (-3.6,-0.7);
 \draw [thick] (-4,-0.7) rectangle (-3.6,-0.1);
 \draw [thick, -] (-4,-0.3) -- (-3.6,-0.3);
 \draw [thick, -] (-3.8,-1.5) -- (-3.8,-2);
 \draw [thick, ->] (-3.8,0.5) -- (-3.8,-0.1);
 \small
 \node at (0,-2.2) {$h_n$};
 \node at (-2.2,-1) {$u_n$};
 \node at (-1,-0.05) {$q$};
 \node at (-1,-1.4) {$z_n$};
 \node at (-4.1,0.25) {${\cal E}_n$};
 \node at (-4.35,-1) {${\cal B}_n$};
 \end{tikzpicture}
 }
\vspace{2mm}
 \caption{Our CR system model corresponding to SU$_n$ for $n=1, \ldots ,\Nu$.} 
 \label{CRStructure}
 \vspace{2mm}
 \end{figure}
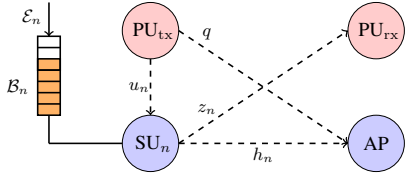
%
%
\begin{figure}[!t]
	\vspace{-0mm}
	\centering
	\scalebox{0.75}{
	\begin{tikzpicture}
	\smaller
	\draw [thick, fill=green!20] (-3.2,0) rectangle (-2,0.8);
	\node at (-2.6,0.4) {Sensing};
	\draw [thick, fill=orange!20] (-2,0) rectangle (-0.8,0.8);
	\node at (-1.4,0.4) {Probing};	
	\draw [thick, fill=gray!20] (-0.8,0) rectangle (1.6,0.8);
	\node at (0.4,0.4) {Data Transmission};
	\draw [thick, fill=green!20] (1.6,0) rectangle (2.8,0.8);
	\node at (2.2,0.4) {Sensing};
	\draw [thick, fill=orange!20] (2.8,0) rectangle (4,0.8);
	\node at (3.4,0.4) {Probing};
	\draw [thick, fill=gray!20] (4,0) rectangle (6.4,0.8);
	\node at (5.2,0.4) {Data Transmission};
	\draw [-] (-3.2,1.05) -- (-3.2,0.75);
	\draw [thick,-] (-3.2,0) -- (-3.2,-0.25);
	\draw [-] (-2,0) -- (-2,-0.25);
	\draw [-] (-0.8,0) -- (-0.8,-0.25);	
	\draw [-] (1.6,1.05) -- (1.6,0.75);	
	
	\draw [thick,-] (1.6,0) -- (1.6,-0.25);
	\draw [-] (2.8,0) -- (2.8,-0.25);
	\draw [-] (4,0) -- (4,-0.25);
	\draw [thick,-] (6.4,0) -- (6.4,-0.25);
	\draw [-] (6.4,1.05) -- (6.4,0.75);
	\draw [<->] (-3.2,-0.15) -- (-2,-0.15);
	\draw [<->] (-2,-0.15) -- (-0.8,-0.15);
	\draw [<->] (-0.8,-0.15) -- (1.6,-0.15);
	\draw [<->] (1.6,-0.15) -- (2.8,-0.15);
	\draw [<->] (2.8,-0.15) -- (4,-0.15);
	\draw [<->] (4,-0.15) -- (6.4,-0.15);
	\draw [<->] (-3.2,0.95) -- (1.6,0.95);
	\draw [<->] (1.6,0.95) -- (6.4,0.95);
	\draw [->] (-3.2,-1) -- (-3.2,-1.6);
	\draw [->] (1.6,-1) -- (1.6,-1.6);
	\draw [->] (6.4,-1) -- (6.4,-1.6);
	\draw [thick, -] (-3.4,-1.6) -- (6.6,-1.6);
	\node at (-2.55,-0.3) {$\Tsen$};
	\node at (-1.35,-0.3) {$\Ttra$};
	\node at (0.45,-0.3) {$\Td$};
	\node at (2.25,-0.3) {$\Tsen$};
	\node at (3.45,-0.3) {$\Ttra$};
	\node at (5.25,-0.3) {$\Td$};
	\node at (-0.8,1.2) {$\Tf$};
	\node at (4,1.2) {$\Tf$};
	\node at (-3.2,-0.7) {${\cal E}^{(t-1)}_{n}$};
	\node at (1.6,-0.7) {${\cal E}^{(t)}_{n}$};
	\node at (6.35,-0.7) {${\cal E}^{(t+1)}_{n}$};
	\node at (-0.9,-1.45) {$t$};
	\node at (4.0,-1.45) {$t+1$};
	\end{tikzpicture}
	}
	\vspace{2mm}	
	\caption{ Slot structure of SUs.} 
	\label{Frame}
	\vspace{0mm}
\end{figure}
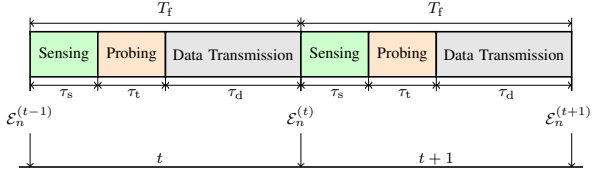
%
%
%
\subsection{Battery and Energy Harvesting Models}

\par We assume that SUs operate under a time-slotted scheme, with slot duration of $\Tf$ seconds, and  they always have data to transmit. Each time slot is indexed by an integer $t$  for $t = 1, 2, \ldots$. The energy harvester  at each SU stores randomly arrived energy in a finite size battery and consumes the stored energy for spectrum sensing, channel probing, and data transmission. Each battery consists of $K$ cells (units) and the amount of energy stored in each unit is equal to $\eu$ Joules. Thus, the battery can store up to $K \eu$ Joules of energy.

\par When $k$ cells of the battery is charged (the amount of stored energy in the battery is $ k \eu$ Joules) we say that the battery {\it is at  state $k$}. Let ${\cal B}_n^{(t)} \in \{0, 1, \ldots, K \} $ denote the discrete random process indicating the battery state of SU$_n$ at the beginning of time slot $t$. We define the probability mass function (pmf) of the discrete random variable ${\cal B}_n^{(t)}$ as $\zeta^{(t)}_{k,n} = \Pr({\cal B}_n^{(t)} = k)$, where $\sum_{k=0}^K \zeta_{k,n}^{(t)} =1$. Note that ${\cal B}_n^{(t)}=0$ and ${\cal B}_n^{(t)}=K$ represent the empty battery and full battery levels, respectively.

\par Let ${\cal E}_n^{(t)}$ denote the randomly arriving  energy packets during time slot $t$ of SU$_n$, where the energy packet measured in Jules is $e_u$ Jules. The discrete random process ${\cal E}_n^{(t)}$  is typically modeled as a sequence of independent and identically distributed (i.i.d.) random variables \cite{FanConf},  regardless of the spectrum occupancy state of \PUtx. We assume that the discrete random variables ${\cal E}_n^{(t)}$'s are i.i.d. over time and independent across sensors. We  model  ${\cal E}_n^{(t)}$ as a Poisson random variable with the pmf $f_{ {\cal E}_n } \!(r) = \Pr({\cal E} = r )= e^{-\rho_n} \rho_n^r / r! $ for $r = 0, 1, \ldots, \infty$,  where $\rho_n$ denotes the {average number of arriving energy packets during one time slot of SU$_n$.}\footnote{We note that $\rho_n$ does not depend on the duration of spectrum sensing phase, since we assume each node is capable of harvesting energy during the entire slot. If we limit harvesting energy to spectrum sensing phase, then  $\rho_n$ would change to $\rho_n \Tsen/\Tf$.} 
Let $\alpha_{h_n}^{(t)}$ be the number of stored (harvested) energy units in the battery at SU$_n$ during time slot $t$. This harvested energy cannot be used during time slot $t$. Since the battery has a finite capacity of $K$ cells, $\alpha_{h_n}^{(t)} \in \{0,1, \ldots ,K \}$. Also, $\alpha_{h_n}^{(t)}$ are i.i.d. over time slots and independent across sensors. Let $f_{\! \alpha_{h_n} } \!\!(r)= \Pr(\alpha_{h_n} =r)$ denote the pmf of $\alpha_{h_n}^{(t)}$.  We  can find the pmf of $\alpha_{h_n}^{(t)}$ in terms of the pmf of ${\cal E}_n^{(t)}$ as the following
%
\begin{equation} 
f_{\! \alpha_{h_n} } \!(r) = \left\{
\begin{array}{ll} 
f_{ {\cal E}_n } \!(r), \quad &\text{if} ~ 0 \leq r \leq K-1\\
\sum_{m = K}^{\infty} f_{ {\cal E}_n } \!(m), \quad &\text{if} ~ r=K.
\end{array}\right. 
\end{equation}
%
%
%
%
\subsection{Slot Structure of SUs}

\par Each time slot consists of three sub-slots (see Fig. \ref{Frame}), corresponding to spectrum sensing phase, channel probing phase, and data transmission phase, with fixed durations of $\Tsen = \Nsen / \fs, \Ttra = \Ntra/ \fs, \Td= \Nd/ \fs$, respectively. Note that $\fs$ is the sampling frequency, $\Nsen$ is the number of collected samples during spectrum sensing phase, $\Ntra$ is the number of training symbols sent during channel probing phase, and $\Nd$ is the number of data symbols sent during data transmission phase. Also, we have $\Tf= \Tsen + \Ttra + \Td$.

\par During {\it spectrum sensing phase}, SU$_n$ senses its pre-assigned single spectrum band to detect \PUtx's activity. We model the \PUtx's activity in each spectrum band as a Bernoulli random variable and we assume the statistics of \PUtx ~are i.i.d.  across $M$ spectrum bands and over time slots. Therefore, we can frame the spectrum sensing problem at SU$_n$ as a binary hypothesis testing problem. Suppose $ \Hl^{(t)}$  and $ \Ho^{(t)}$  represent the binary hypotheses of \PUtx ~being active and inactive in time slot $t$, respectively, with prior probabilities $\Pr \{ \Hl^ {(t)} \} \! = \!  \pi_1$ and $\Pr \{ \Ho^{(t)}\} \! =\! \pi_0$.  SU$_n$ applies a binary detection rule to decide whether or not \PUtx ~is active in its pre-assigned band.  Let $\Hhaton$ and $\Hhatln$, with probabilities $\piohn = \Pr \{ \Hhaton \}$ and $\pilhn = \Pr \{ \Hhatln \}$ denote the SU$_n$ detector outcome, i.e., the detector finds \PUtx ~active and inactive (the result of spectrum sensing is busy or idle), respectively. The accuracy of this binary detector is characterized by its false alarm and detection probabilities. The details of the binary detector are presented in Section \ref{EnergyDetector}.

\par Depending on the outcome its of spectrum sensing, SU$_n$ stays in {spectrum sensing phase}  or enters {\it{channel probing phase}}. In this phase,  SU$_n$ sends $\Ntra$ training symbols with fixed symbol power $\Ptra = \alphatra \eu / \Ttra$, to enable channel estimation at the AP, where $\alphatra$ is the number of consumed cells of energy for channel probing\footnote{For ease of presentation, we assume that circuit power (energy) consumption is negligible in comparison to the consumed energy for channel probing and data transmission. Otherwise, it can easily be incorporated into the system model.}. We assume that the battery always has $\alphatra$ units of stored energy for channel probing. Let $h_n^{(t)}$ denote the SU$_n$-AP fading coefficient in time slot $t$ and  $g_n^{(t)} = |h_n^{(t)}|^2$ be the corresponding channel power gain. Using the received signals corresponding to the training symbols, the AP estimates $\widehat{h}_n^{(t )}$ and lets $\ghatn^{(t)}=|\widehat{h}_n^{(t )}|^2$ and shares this value with SU$_n$ via a feedback channel. 
Next, SU$_n$ enters {\it data transmission phase}. During this phase, SU$_n$ sends $ \Nd$ Gaussian data symbols with adaptive symbol power according to its battery state and the received information via the feedback channel about SU$_n$--AP link. If  the battery is at state $k$, then SU$_n$ allocates $\alpha_{k,n}$ cells of stored energy for each data symbol transmission, implying that the adaptive symbol power is $P_{k,n}^{(t)} = \alpha_{k,n}^ {(t)} \pu$, where  $\pu = \eu / \Td$. Note that since  $\alpha_{k,n}^{(t)}$  is discrete, $P_{k,n}^{(t)}$  is discrete. The details of the choice of $\alpha_{k,n}^ {(t)} $ according to the battery state $k$ and the feedback information $\ghatn$ are given in Section \ref{Transmission_Model} and the details of channel estimation are explained in Section \ref{Ch_Training}.
%
%
\subsection{Transmission Model and Battery Dynamics} \label{Transmission_Model}

\par As we said, we assume that SU$_n$ adapts its transmit energy per data symbol (power) according to its battery state $k$ and the received information via the feedback channel about its channel power gain $\ghatn$. In particular, we choose a power adaptation strategy that mimics the behavior of the rate-optimal power adaptation scheme with respect to the channel power gain \cite{J2}, i.e., when $\ghatn$ is smaller than a cut-off threshold $\theta_n$ (to be optimized), the transmit {energy} is zero, and when $\ghatn$  exceeds $\theta_n$, the transmit {energy} increases monotonically as $\ghatn$ increases until it reaches its maximum value of {$\lfloor k \Omega_n \rfloor - \alphatra$}, where $\Omega_n \in [0,1]$ (to be optimized), and $\lfloor \cdot \rfloor$ denotes the floor function.  Mathematically, we express $\alpha_{k,n}^{(t)}$ for SU$_n$ as the following
%
%
\begin{subequations} \label{alphaBAR}
	\begin{align}
	&\alpha_{k,n}^{(t)} = \max \big \{ \overline{\alpha}_{k,n}^{(t)} \,,\, 0 \big \}, ~~~\text{for} ~ k  =  0, 1, \ldots ,K, \label{alphaBAR_a}\\
	&\overline{\alpha}_{k,n}^{(t)} =  \Big \lfloor \Omega_n \,k \,\Big ( 1 - { \theta_n \over \ghatn ^{(t)} } \Big )^+  \Big \rfloor - \alphatra, \label{alphaBAR_b}
	\end{align}
\end{subequations}			
%
%
\noindent where $(x ) ^{+} = \max\{x, 0\}$. The parameters $\Omega_n$ and $\theta_n$ in \eqref{alphaBAR} play key roles in balancing the energy harvesting and the energy consumption. Given $\theta_n$, when  $\Omega_n$ is large {(or given $\Omega_n$, when $\theta_n$ is small)}, such that the rate of energy consumption is greater than the rate of energy harvesting, SU$_n$ may stop functioning, due to energy outage. On the other hand, given $\theta_n$, when $\Omega_n$ is small {(or given $\Omega_n$, when $\theta_n$ is large)}, SU$_n$ may fail to utilize the excess energy, due to energy overflow, and the data transmission would become limited in each slot. Note that $\overline{\alpha}_{k,n}^{(t)}$ in \eqref{alphaBAR} ensures that the battery always has $\alphatra$ units of stored energy for channel probing. Furthermore, the transmission policy in \eqref{alphaBAR} satisfies the  causality constraint of the battery. The causality constraint restrains the energy corresponding to symbol transmit power to be less than the available stored energy in the battery, i.e., $\alpha^{}_{k,n} \leq  { k } - \alphatra$. Note that $\alpha_{k,n}$ is discrete random variable and $\alpha_{k,n} \in \{ 0, 1, \ldots , K\}$. Let	$\psi_{i,k,n}^ \varepsilon = \Pr  ( \alpha_{k,n} = i | \Heps )$  denote the pmf of $\alpha_{k,n}$ given $\Heps, \varepsilon =0, 1$. We have
%
%
\begin{align} \label{Prob_alpha}
	\psi_{i,k,n}^ \varepsilon = \left\{
	\begin{array}{ll} 
		1 ,  & \text{if}~~ 0 \leq k \leq \alphatra, ~i=0  \\
		0, &  \text{if}~~ 0 \leq k \leq \alphatra, ~i \neq 0  \\
		Y_{k,n}, & \text{if}~~ k \geq \alphatra \! + \!1, ~i=0 \\
		Q_{i,k,n}, & \text{if}~~ k \geq \alphatra \! + \!1, ~ 1 \leq i \leq  \lfloor k \Omega_n \! \rfloor \! - \! \alphatra \\
		0, & \text{if}~~ k \geq \alphatra \! + \!1, ~ i \geq  \lfloor k \Omega_n \! \rfloor \! - \! \alphatra + 1
	\end{array}\right.   
\end{align}
%
in which
%
\begin{subequations}
	\begin{align}
	&Q_{i,k,n} =  F^ \varepsilon_{\ghatn} \big ( c_{i,k,n} \big ) - F^ \varepsilon_{\ghatn } \big ( a_{i,k,n } \big ) \\
	&Y_{k,n} = F^ \varepsilon_{\ghatn} (\theta_n)  + \sum_{m=1}^{ \min (\lfloor k \Omega_n \! \rfloor , \alphatra ) } Q_{m-\alphatra,k,n } \\
	a_{i,k,n} = &\frac{\theta_n k \Omega_n}{k \Omega_n \! - \! \alphatra \! - \! i}, ~~~~~ c_{i,k,n} = {\theta_n k \Omega_n \over k \Omega_n  \!- \! \alphatra \! - \!i \!- \!1}, \label{a_c}
	\end{align}
\end{subequations}			
%
%
%
where $F^ \varepsilon_{\ghatn} (x) = F_{\ghatn} (x| \Heps)$ is the cumulative distribution function (CDF) of $\ghatn$ given $\Heps$. Note that if $c_{i,k,n} < 0$, we set $c_{i,k,n} = +\infty$.

\par The battery state at the beginning of time slot $t+1$ depends on the battery state at the beginning of time slot $t$, the harvested energy during time slot $t$, the transmission symbol, as well as $\alphatra$. In particular, if at time slot $t$, SU$_n$ senses its spectrum band to be idle, the state of its battery at the beginning of slot $t + 1$ is	
%
\begin{equation} \label{Batery_state0}
{\cal B}_n^{(t+1)} = \min \left \{ \Big ( {\cal B}_n^{(t)}  - \alphatra  - \alpha_{k,n}^{(t)} + \alpha_{h_n}^{(t)} \Big )^+ , \; K  \right \}.
\end{equation}
%
On the other hand if at time slot $t$, SU$_n$ senses its spectrum band to be busy, the state of its battery at the beginning of slot $t+1$ is
%
\begin{equation} \label{Batery_state1}
{\cal B}_n^{(t+1)} = \min \left \{ \Big ( {\cal B}_n^{(t)} + \alpha_{h_n}^{(t)} \Big )^+ , \; K  \right \},
\end{equation}
%
since $\alpha_{k,n}^{(t)} =0$. Considering the dynamic battery state model in \eqref{Batery_state0} and \eqref{Batery_state1} we note that, conditioned on $\alpha_{h_n}^{(t)}$ and $\alpha_{k,n}^{(t)}$ the value of ${\cal B}_n^{(t+1)}$ only depends on the value of ${\cal B}_n^{(t)}$ (and not the battery states of time slots before $t$). Hence, the battery state random process ${\cal B}_n^{(t)}$ can be modeled as a Markov chain. Let the probability vector of battery state in time slot $t$ be $\boldsymbol{\zeta}_n^{(t)} = [\zeta_{1,n}^{(t)}, \ldots , \zeta_{K, n}^{(t)}]^T$. Note that the probability $\zeta_{k,n}^{(t)}$ depends on the battery state at slot $t-1$,  the number of battery units filled by the harvested energy during slot $t-1$,  the probability of spectrum band sensed idle, and, the number of energy units allocated for data transmission at slot $t-1$ when the spectrum band is sensed idle, i.e., $\zeta_{k,n}^{(t)}$ depends on  ${\cal B}_n ^{(t-1)}, \alpha_{h_n }^{(t-1)}, {\piohn }, \alpha^ {(t-1)}_{k,n}$, respectively. Assuming the Markov chain is time-homogeneous, we let $\boldsymbol{\Phi}_n$ denote the $(K+1) \times (K+1)$ transition probability matrix of this chain with its $(i,j)$-th entry $\phi^n_{i,j}  = \Pr({\cal B}_n^{(t)} =i | {\cal B}_n^{(t-1)} = j)$ given in \eqref{zeta} where $F_{\! \alpha_{h_n}} \!\!(\cdot)$ is the commutative distribution function (CDF) of $\alpha_{h_n}$. We have
%
%
%
\begin{figure*}
	\begin{subequations}\label{zeta}
		\begin{align}
			&\phi^n_{0,j}  = \sum_{l=0}^{K}  \Big [ \psi_{l,j,n}^0  \piohn F_{\! \alpha_{h_n}} \!\! \big ( \alphatra \! + \!  l \! - \!j \big )  \Big ] + \pilhn F_{\! \alpha_{h_n}} \!\! \big ( - j \big )  \label{zeta0}\\
			&\phi^n_{K,j} = \sum_{l=0 }^{K} \Big [ \psi_{l,j,n}^0  \piohn \Big ( 1 \! - \! F_{\! \alpha_{h_n}} \!\! \big ( \alphatra \! + \!  l \! + \! K\! - \!j \big ) \Big )  \Big ] + \pilhn \Big ( 1 \! - \! F_{\! \alpha_{h_n}} \!\! \big ( K \! - \! j \big ) \Big ) \label{zetaK} \\
			&\phi^n_{i,j} = \sum_{l=0}^{K}  \Big [ \psi_{l,j,n}^0 \piohn f_{\! \alpha_{h_n}} \!\! \big ( \alphatra \! + \!  l \! + \! i \! - \!j \big ) \Big ]	 + \pilhn f_{\! \alpha_{h_n}} \!\! \big (  i \! - \! j \big )  \!,  ~~~~\text{for} ~i=1, \ldots ,K-1 \label{zeta1K}
		\end{align}
	\end{subequations}
	\hrulefill
\end{figure*}
%
%
%
\vspace{-1mm}
\begin{equation} \label{MarkPhi}
\boldsymbol{\zeta}_n^{(t+1)} = \boldsymbol{\Phi}_n  \,\boldsymbol{ \zeta}_n^{(t)}.
\end{equation}
%
Since the Markov chain characterized by the transition
probability matrix $\boldsymbol{\Phi}_n$ is irreducible and aperiodic, there exists a unique steady-state distribution, regardless of the initial state \cite{QueuingFundamentals}. Let $\boldsymbol{\zeta}_n = \lim_{t \to \infty} \boldsymbol{\zeta}^{ (t)}_n$ be the unique  steady-state probability vector. This vector satisfies the following equations
\vspace{-2mm}
\begin{subequations}
	\begin{align}
	\boldsymbol{\zeta}_n & = \boldsymbol{\Phi}_n  \,\boldsymbol{ \zeta}_n \label{MarkovEq}, \\
	\boldsymbol{\zeta}_n^T  \boldsymbol{1} & = \sum_{k=1}^{K} \zeta_{k,n} = 1,
	\end{align}
\end{subequations}
%
%
\noindent where $\boldsymbol{1}$ is an all-ones vector, i.e.,  $\boldsymbol{ \zeta}_n $ is the normalized eigenvector corresponding to the unit eigenvalue of $\boldsymbol{\Phi}_n$, such that the entries of $\boldsymbol{\zeta}_n$ sums up to one.
The closed-form expression for $\boldsymbol{\zeta}_n$ is \cite{Zhang_Full_Duplex}
%
\begin{equation} \label{zeta_fin}
\boldsymbol{\zeta}_n = \big (\boldsymbol{\Phi}_n - \boldsymbol{I} + \boldsymbol{B} \big )^{-1} \,\boldsymbol{1},
\end{equation}
%
where $\boldsymbol{B}$ is an all-ones matrix and $\boldsymbol{I}$ is the identity matrix. From this point forward, we assume that the battery is at its steady-state and we drop the superscript $t$. 
\par To illustrate our transmission model in \eqref{alphaBAR} we consider the following simple numerical example. Assuming that the battery has $K=7$ cells, Fig. \ref{alphaTr} shows an example of $\alpha_{k,n}$ for our CR system for two sets of $\{ \Omega_{n}, \theta_{n} \}$ given as  $\Omega_n^{(a)} = 0.75 , \theta_n^{(a)} = 0.02$  and $\Omega_n^{(b)} = 0.95, \theta_n^{(b)} = 0.05 $. The corresponding transition probability matrices are given in the following 
%
%
\begin{subequations}
\begin{equation*} \label{Matrix_a}
\boldsymbol{\Phi}_n^{(a)} \! = \!
\small 
\begin{pmatrix}
0.42&0.29&0.17&0.08&0.02&0&0&0\\
0.12&0.13&0.12&0.09&0.05&0.02&0&0\\
0.19&0.12&0.13&0.12&0.09&0.05&0.02&0\\
0.07&0.19&0.12&0.13&0.12&0.09&0.05&0.02\\
0.05&0.07&0.19&0.12&0.13&0.12&0.09&0.05\\
0.05&0.05&0.07&0.19&0.12&0.13&0.12&0.09\\
0.04&0.05&0.05&0.07&0.19&0.12&0.13&0.12\\
0.06&0.1&0.15&0.2&0.27&0.46&0.58&0.71\\
\end{pmatrix},
\end{equation*}
\begin{equation*} \label{Matrix_b}
\boldsymbol{\Phi}_n^{(b)}  \! = \!
\small 
\begin{pmatrix}
0.51&0.4&0.28&0.16&0.08&0.02&0&0\\
0.17&0.11&0.12&0.12&0.09&0.05&0.02&0\\
0.06&0.17&0.11&0.12&0.12&0.09&0.05&0.02\\
0.05&0.06&0.17&0.11&0.12&0.12&0.09&0.05\\
0.05&0.05&0.06&0.17&0.11&0.12&0.12&0.09\\
0.05&0.05&0.05&0.06&0.17&0.11&0.12&0.12\\
0.04&0.05&0.05&0.05&0.06&0.17&0.11&0.12\\
0.06&0.11&0.16&0.21&0.26&0.32&0.49&0.6\\
\end{pmatrix}.
\end{equation*}
\end{subequations}
%
%
\begin{figure}[!t]
	\vspace{-0mm}
	\centering
	\begin{subfigure}[b]{0.5\textwidth}                
		\centering     		
		\includegraphics[width=55mm]{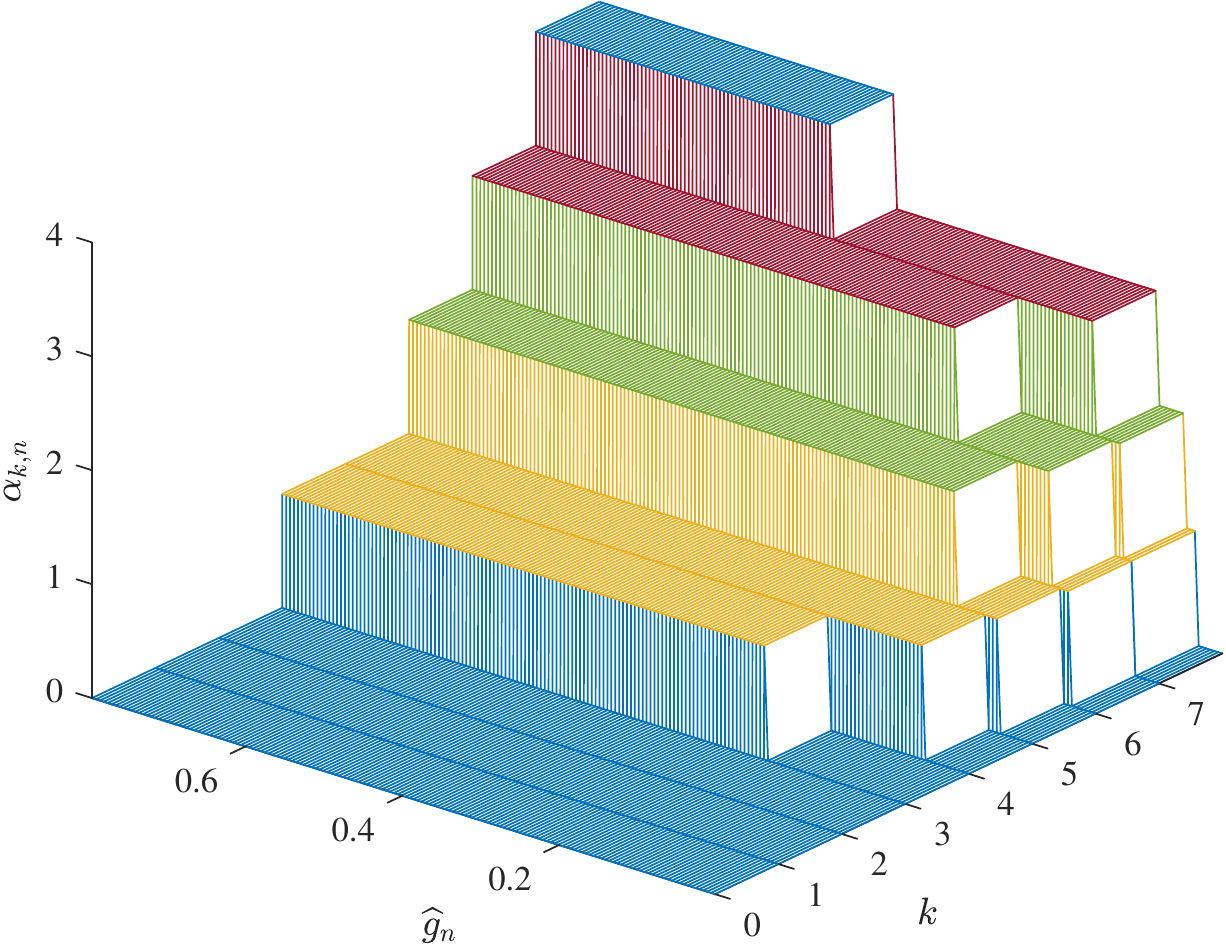}
		\caption{} 
		\label{}      
	\end{subfigure}\\
	\vspace{2mm}
	\begin{subfigure}[b]{0.5\textwidth}
		\centering	              		
		\includegraphics[width=55mm]{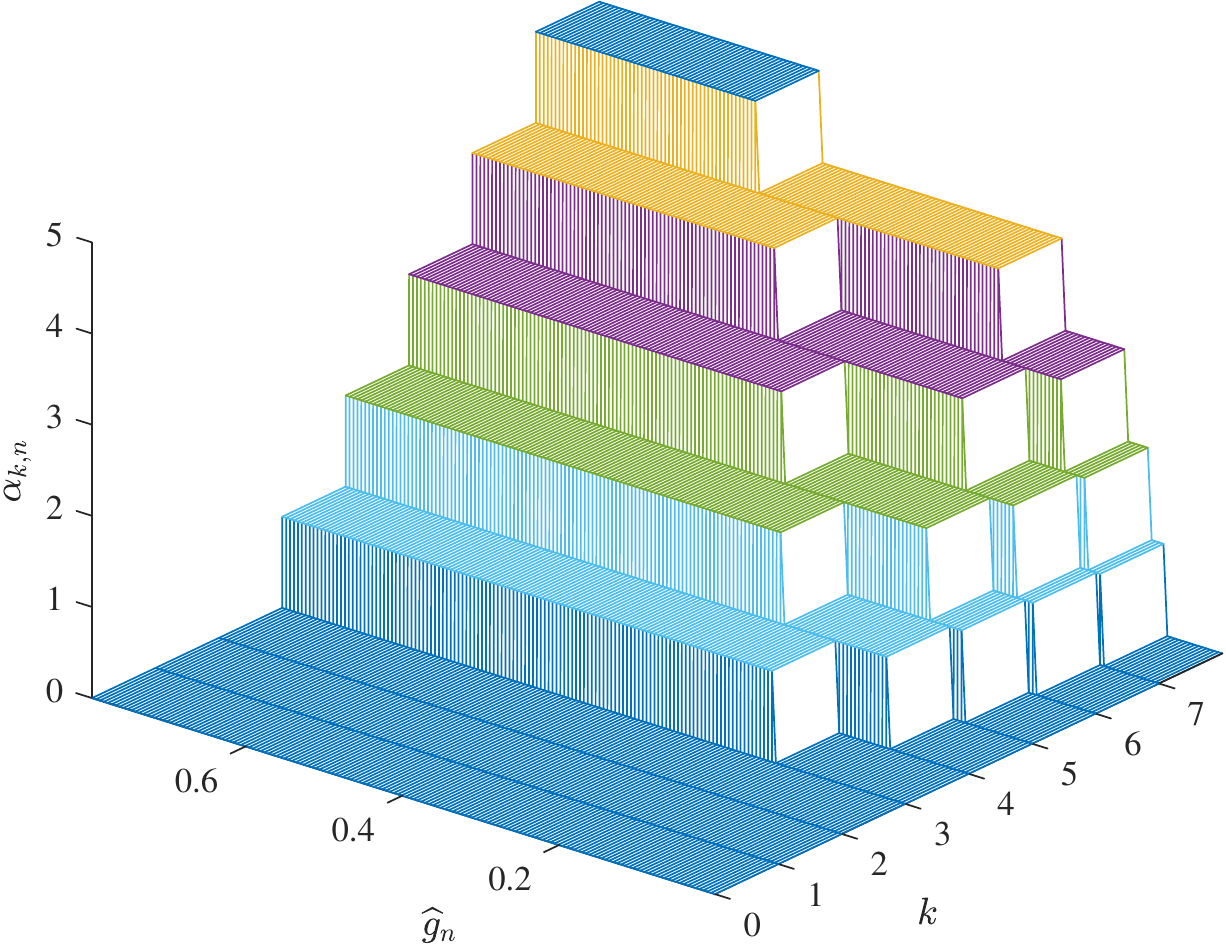}
		\caption{} 
		\label{}     
	\end{subfigure} \\
	\vspace{0mm}
	\caption{This example shows how many energy units ($\alpha_{k,n}$) SU$_n$ spends for data transmission, given its battery state and the received information about its channel gain via feedback link. (a) $\Omega_n^{(a)} = 0.75 , \theta_n^{(a)} = 0.02$, (b) $\Omega_n^{(b)} = 0.95, \theta_n^{(b)} = 0.05 $.} 
	\label{alphaTr}
\end{figure}
%
%
\par Our goal is to find the transmission parameters $\{\Omega_n,\theta_n\}$ in \eqref{alphaBAR_b} for all SUs such that the sum-rate of our CR network is maximized, subject to a constraint on the average interference power that collective SUs can impose on \PUrx.
We assume that this optimization problem is solved {\it offline} at AP, given the statistical information of fading channels, the number of samples collected during spectrum sensing phase $\Nsen$, the number of training symbols sent during channel probing phase $\Ntra$, and power of training symbols $\Ptra$. The solutions to this optimization problem, i.e., the optimal set $\{\Omega_{n}, \theta_{n} \}_{n=1} ^{\Nu}$ is available {\it a priori} at the AP and SUs, to be utilized for adapting symbol power during data transmission phase. The idea of offline power allocation optimization with a limited feedback channel has been used before for distributed detection systems in wireless sensor networks \cite{Evans1}. In the following sections, we describe how SUs operate during {spectrum sensing phase}, {channel probing phase}, and {data transmission phase}. For the readers' convenience, we have collected the most commonly used symbols in Table \ref{table1}.
%
%
\begin{table}[h!]
	\begin{center}
		\caption{Most commonly used symbols.}
		\label{table1}
		\begin{footnotesize}
			\begin{tabular}{l|l} 
			\!\!\!\!\! \textbf{Symbol} & \!\!\!\! \textbf{Description} \\
			\hline \hline
			\!\!\!\!\! $\Nsen$ & \!\!\!\! Number of collected samples during {\it spectrum sensing phase}\\
			\!\!\!\!\! $\Ntra$ & \!\!\!\! Number of training symbols during {\it channel probing phase} \\
			\!\!\!\!\! $\Nd$ & \!\!\!\! Number of  data symbols during {\it data transmission phase} \\		
			\!\!\!\!\! $\Ptra$ & \!\!\!\! Power of training symbols \\      
			\!\!\!\!\! $h_n, \widehat{h}_n, \widetilde{h}_n$ \!\!\!\!\!\! & \!\!\!\! Fading coefficient of SU$_n$--AP link, LMMSE channel \\ &  \!\!\!\! estimate, and its corresponding estimation error \\ 
			\!\!\!\!\! $\gamn, \gamhatn, \gamtiln$ \!\!\!\!\!\!\! &  \!\!\!\! Variances of $h_n, \widehat{h}_n, \widetilde{h}_n$  \\           
			\!\!\!\!\! $\pi_0, \pi_1$ & \!\!\!\! Prior probabilities of $\Ho$  and $\Hl$ \\ 
			\!\!\!\!\! $\piohn, \pilhn$ \!\!\!\!\!\!\! &  \!\!\!\! Probabilities of spectrum bands being sensed idle or busy \\       
			\!\!\!\!\! $\zeta_{k,n}$ & \!\!\!\! Probability of SU$_n$  battery being at state $k$\\
			\!\!\!\!\! $u_n$ & \!\!\!\! Fading coefficient  of \PUtx--SU$_n$ link with variance $\delta_{ u_n}$\\
			\!\!\!\!\! $q$ & \!\!\!\! Fading coefficient of \PUtx--AP link with variance $\delta_{q}$\\
			\!\!\!\!\! $z_n$ & \!\!\!\! Fading coefficient  of SU$_n$--\PUrx ~link with variance $\delta_{ z_n}$
			\end{tabular}
		\end{footnotesize}   
	\end{center}   
	\vspace{-5mm}  
\end{table}
%
%
%
\section{Spectrum Sensing Phase} \label{EnergyDetector}

\par In order to access its spectrum band, SU$_n$ first needs to sense its band during spectrum sensing phase, to determine whether it is busy or idle (see Fig. \ref{Frame}). We formulate the spectrum sensing at SU$_n$ as a binary hypothesis testing problem, where the received signal at SU$_n$  can be written as:

%
\begin{equation}\label{E43}
\begin{array}{lll}
\Ho  & : & y_n[m] = w_n[m],\\
\Hl & : & y_n[m] = u_n[m]  \,p[m] + w_n[m], \\
\end{array}%
\end{equation}
%
%
for $m = 1, ..., \Nsen$, where $p[m]$ is the transmit signal of \PUtx, $w_n[m] \sim {\cal CN} (0, \Sigmawn^2)$ is the additive white Gaussian noise (AWGN) at SU$_n$ and $u_n[m]$ is the  fading coefficient corresponding to \PUtx--SU$_n$ channel. The two hypotheses $\Ho$ and $\Hl$ with probabilities $\pi_0$ and $\pi_1 = 1 - \pi_0$ denote the spectrum is truly idle and truly busy, respectively. We assume that $\pi_0$ and $\pi_1$ are known to SUs based on long-term spectrum measurements. For spectrum sensing we consider energy detector, where the decision statistics at SU$_n$ is $Z_n= {1 \over \Nsen} \sum_{m=1 }^{\Nsen} |y_n[m]|^2 $. The accuracy of this detector is characterized by its false alarm probability $P_{{\rm fa}_n}= \Pr( \Hhatln | \Ho ) = \Pr( Z_n > \xi_n | \Ho)$ and detection probability $P_{{\rm d}_n}= \Pr(\Hhatln |\Hl) = \Pr( Z_n > \xi_n | \Hl)$, where $\xi_n$ is the local decision threshold. For large $\Nsen$, we can invoke central limit theorem and approximate the cumulative distribution function (CDF) of $Z_n$ as Guassian. Hence, $P_{{\rm fa}_n}$ and $P_{{\rm d}_n}$ can be expressed in terms of $Q$ function as below \cite{GlobalSIP}
\vspace{-2mm}
\begin{subequations} \label{Pd_Pf}
	\begin{align}
	P_{{\rm fa}_n}  =  &\, Q \bigg (\! \Big( \! \frac{\xi_n}{\Sigmawn^2} -\!1 \! \Big)  \sqrt{\!\Nsen } \!\bigg ), \label{P_dn}\\
	P_{{\rm d}_n}  =  &\, Q \bigg (\!\ \Big( \! \frac{\xi_n}{\Sigmawn^2 } - \nu_n\!-\!1 \! \Big)  \sqrt{ \frac{\!\Nsen }{\!\!2\nu_n \!+ \!1}} \bigg ), \label{P_fn}
	\end{align}
\end{subequations} 
%
where $\nu_n = P_{\rm{p}} \delta_{u_n} /\Sigmawn^2 $ and $P_{\rm{p}}$ is the average transmit power of \PUtx. For a given value of $P_{{\rm d}_n} = \overline{P}_{\rm d}$, the false alarm probability can be written as
%
\begin{equation}\label{}
P_{{\rm fa}_n} = Q \left( \sqrt{2\nu_n+1} Q^{-1} \big (\overline{P}_{ \rm{d}} \big ) +\nu_n \sqrt{\Tsen \fs}  \right ).
\vspace{-0mm}
\end{equation}
%
The probabilities $\piohn$ and $\pilhn$, are related to $P_{{\rm d}_n}$ and $P_{{\rm fa}_n}$. In particular, we have $\piohn  = \betaon + \betaln$ and $\pilhn = 1 - \piohn$ where
\vspace{-2mm}
\begin{subequations} \label{Beta}
	\begin{align}
	\betaon = \Pr \{ \Ho , \Hhaton \}  = \pi_0 (1  -  P_{{\rm fa}_n} ), \\
	\betaln  =  \Pr \{ \Hl ,\Hhaton\}  =  \pi_1 (1  - P_{{\rm d}_n}).
	\end{align}
\end{subequations}
%
%
\section{Channel Probing Phase} \label{Ch_Training}
		
\par Depending on the outcome of its spectrum sensing, SU$_n$  either stays in spectrum sensing phase (i.e., remains silent in the remaining of time slot) if its band is sensed busy (the detector outcome is $\Hhatln$), or it enters channel probing phase if its band is sensed idle (the detector outcome is $\Hhaton$). During channel probing phase, we assume SU$_n$ sends training vector $\boldsymbol {x}_t = \sqrt{\Ptra} \,\boldsymbol{1}$, where $\boldsymbol{1}$ is an $\Ntra \times 1$  all-ones vector to enable channel estimation at the AP. Let vector $\boldsymbol{s}_n = \big [s_n(1) , \ldots, s_n(\Ntra) \big ]^T$ denote the discrete-time representation of received training symbols at the AP from SU$_n$. Assuming the fading coefficient $h_n$ corresponding to SU$_n$--AP channel is unchanged during the entire time slot, we have 
%
\begin{equation} 
\begin{aligned}
\Ho, \Hhaton\!: & ~~~s_n[m] =  h_n \sqrt{\Ptra} + { v_n[m]}, \\
\Hl, \Hhaton\!: & ~~~s_n[m] = h_n \sqrt{\Ptra} + { q[m]}  \,p[m] + { v_n [m]},
\end{aligned}
\end{equation}
%
for $m=1, \ldots, \Ntra$, $v_n[m] \sim {\cal CN} \big (0,\Sigmavn^2 \big )$ is the AWGN at the AP,  and $q[m]$ is the fading coefficient corresponding to \PUtx--AP channel. The linear minimum mean square error (LMMSE) estimate of $h_n$ given $\Hhaton$ is \cite{kay93, J2}
%
\begin{subequations} \label{LMMSE_eq}
	\begin{align}
	\widehat{h}_n =\,& C_{h_n \boldsymbol{s}_n} C^{-1}_{\boldsymbol{s}_n} \,\boldsymbol{s}_n, \label{Xhat} \\
	C_{h_n \boldsymbol{s}_n} =\,& \mathbb{E} \{h_n \boldsymbol{s} ^H_n | \Hhaton \} \! = \!\gamn \sqrt{\Ptra}  \,\boldsymbol{1},\\
	C_{\boldsymbol{s}_n } =\,& \mathbb{E} \big \{\boldsymbol{s}_n \boldsymbol{s}^H_n | \Hhaton \big \} \! = \! \omegaon \,\mathbb{E} \big \{\boldsymbol{s }_n \boldsymbol{s}^H_n | \Ho, \Hhaton \big \} \nonumber \\ 
	& \qquad \qquad \qquad~ +  \omegaln \,\mathbb{E} \big \{\boldsymbol{s}_n \boldsymbol{s}^H_n | \Hl, \Hhaton \big \} ,
	\end{align}	
\vspace{-2mm}
\end{subequations}
%
where
%
\begin{subequations} \label{Omega_eq}
	\begin{align}
	\omegaon = \Pr \{ \Ho| \Hhaton \}  = {\pi_0 (1 -  {P}_{{\rm fa}_n} ) \over \piohn} = {\betaon \over \piohn}, \\
	\omegaln  =  \Pr \{\Hl | \Hhaton \}  = {\pi_1 ( 1 - {P}_{{\rm d}_n} ) \over \piohn} = {\betaln \over \piohn},
	\end{align}
\vspace{-2mm}
\end{subequations}
%
and
%
\begin{subequations} 
	\begin{align}
		\mathbb{E} \big \{\boldsymbol{s }_n \boldsymbol{s}^H_n | \Ho, \Hhaton \big \} = & \,(\gamhatn^{0} \Ptra + \Sigmavn^2 ) \,\boldsymbol{I}, \\
		\mathbb{E} \big \{\boldsymbol{s }_n \boldsymbol{s}^H_n | \Hl, \Hhaton \big \} = & \,(\gamhatn^{1} \Ptra + \Sigmavn^2 + \Sigmap^2 ) \,\boldsymbol{I}.
	\end{align}
\vspace{-1mm}
\end{subequations}
%
After substituting \eqref{Omega_eq} into \eqref{LMMSE_eq}, $\widehat{h}_n$ reduces to 
\vspace{-1mm}
\begin{equation}\label{chi_hat}
\widehat{h}_n = \frac{\gamn \sqrt{\Ptra}  }{ \gamn  \Ptra \Ntra + { \Sigmavn^2 } + \omegaln { \Sigmap^2 } } \sum_{m=1}^{\Ntra } s_n[m],
\end{equation}
%
%
\noindent where {$\Sigmap^2 = \Pp \delta_{q}$}. The estimation error is $\widetilde{h}_n = h_n - \widehat{h}_n$, where $\widehat{h}_n$ and $\widetilde{ h}_n$ are orthogonal random variables \cite{kay93}, and $\widehat{h}_n$ and $\widetilde{h}_n$ are zero mean. Approximating  $q[m] p[m]$ as a zero-mean Gaussian random variable with variance ${ \Sigmap^2 }$, we find that the estimate $\widehat{h}_n$ given $\Hhaton$  is distributed as a Gaussian mixture random variable \cite{J2}. Let $\gamhatn$ and $\gamtiln$, represent the variances of $\widehat{h}_n$ and $\widetilde{h}_n$, respectively. Also, Let $\gamhatn^0$ and $\gamhatn^1$ represent the variances of $\widehat{h}_n$ under $\{\Ho , \Hhaton\}$ and $\{\Hl , \Hhaton  \}$, respectively. We have
%
\begin{subequations}\label{var_H}
	\begin{equation}
	\gamhatn^{0} \! =\! \mathbb{VAR} \{\widehat{h}_{n} | \Ho ,{ \Hhaton}  \} \! =\!  \frac{ \gamn^2 \Ptra \Ntra \big (\gamn \Ptra \Ntra \! + \! \Sigmavn^2 \big ) }{\big (\gamn \Ptra \Ntra \! + \! \Sigmavn^2 \! + \! \omegaln { \Sigmap^2 } \big )^2}, \label{var_H0}
	\end{equation}
	\begin{equation}
	\gamhatn^{1} \!=\!  \mathbb{VAR} \{\widehat{h}_{n} | \Hl , { \Hhaton}  \} \!=\! \frac{ \gamn^2 \Ptra \Ntra \big ( \gamn \Ptra \Ntra \! + \! { \Sigmavn^2 } \! + \! \Sigmap^2  \big ) }{\big ( \gamn  \Ptra \Ntra + { \Sigmavn^2 }  \! + \! \omegaln { \Sigmap^2 } \big )^2 }. \label{var_H1}
	\end{equation}
\end{subequations}
%
%
\noindent Therefore, $\gamhatn = { \omegaon } \gamhatn^0   + { \omegaln } \gamhatn^1$. Also, let $\gamtiln^0$ and $\gamtiln^1 $  indicate the variances of $\widetilde{h}_n$ under $\{\Ho , \Hhaton\}$ and $\{\Hl , \Hhaton  \}$, respectively. We have
%
\begin{subequations} \label{ErrVariance}
	\begin{align}
	\gamtiln^0 = & \,\mathbb{VAR} \{\widetilde{h}_n| \Ho, \Hhaton \}= \gamn- \gamhatn^0,  \\
	\gamtiln^1 = & \,\mathbb{VAR} \{\widetilde{h}_n| \Hl, \Hhaton  \}= \gamn- \gamhatn^1.
	\end{align}
\end{subequations}
%
Hence,  $\gamtiln = { \omegaon } \gamtiln^0 + { \omegaln } \gamtiln^1$. For ideal spectrum sensing, we get $\omegaon = 1$ and $\omegaln = 0$ and $\widehat{h}_n$ becomes Gaussian. Let $F^\varepsilon_{\ghatn}(x)$ denote the CDF of $\ghatn$ under $\{\Heps , \Hhaton  \}$ for $\varepsilon=0, 1$. Note that under $\{\Heps , \Hhaton \}$ for $\varepsilon=0, 1$,  $\widehat{h}_n$ is zero mean complex Gaussian. Hence, under  $\{\Heps , \Hhaton \}$ for $\varepsilon=0, 1$, $\ghatn$ is  an exponential random variable with mean $\gamhatn^{\varepsilon}$ and CDF
%
\begin{equation}
F^\varepsilon_{\ghatn}(x) = 1- e^{\frac{-x}{ \gamhatn^{ \varepsilon}}  }.
\end{equation}
%
\noindent The CDF of $\ghatn$, denoted as  $F^\varepsilon_{ \ghatn}(x)$,  can be expressed in terms of  $F^0_{\ghatn}(x)$ and  $F^1_{\ghatn}(x)$ as the following:
%
\begin{equation}
F_{\ghatn }(x) = \omegaon \,F^0_{\ghatn}(x) + \omegaln \, F^1_{\ghatn}(x).
\end{equation}
%
After channel estimation, the AP feeds back the channel gains $\ghatn = |\widehat{h}_n |^2$ over a feedback link to SU$_n$. 
%
%
%
\section{Data Transmission Phase} \label{Phase3}

\par After channel probing phase,  SU$_n$ enters this phase. We note that entering this phase is only possible, if in spectrum sensing phase the outcome of the binary detector is $\Hhaton$. During  this phase, SU$_n$ sends Gaussian data symbols to the AP, while it adapts its transmission power according to  information provided by the AP through the feedback channel about SU$_n$--AP link as well as its battery state. 
In particular, SU$_n$ transmits $\Nd$ zero-mean i.i.d. complex Gaussian symbols $x_{n}[m]$ for $m = 1, \dots, \Nd$ with power $P_{k,n} = \alpha_{k,n} \,\pu$, when the battery is at state $k$ and  $\alpha_{k,n}$ is given in \eqref{alphaBAR}. Let $s_n [m]$ denote the discrete-time representation of received signal at the AP from SU$_n$. Due to error in spectrum sensing, we need to distinguish the signal model for $s_n [m]$ under $\Ho$ and $\Hl$. We have
\vspace{-2mm}
\begin{equation} \label{DataModel}
\begin{aligned}
\Ho, \Hhaton\!: & ~~~s_n[m] =  h_n x_n[m] + { v_n[m]}, \\
\Hl, \Hhaton\!: & ~~~s_n[m] = h_n x_n[m] + { q[m]}  \,p[m] + { v_n [m]}.
\end{aligned}
\end{equation}
%
\noindent Substituting $h_n = \widehat{h}_n + \widetilde{h}_n$ in \eqref{DataModel}, we reach at
%
%
\begin{equation} \label{etaEquation}
\begin{aligned}
\!\Ho ,\! \Hhaton\!\! : & ~~s_n\! [m] \! = \! \widehat{h}_n x_n\! [m] \! + \! \overbracket{ \widetilde{h}_n x_n\! [m] \!+ \! v_n\! [m] }^{\text{new noise} ~\eta_{n,0}[m]}, \\
\!\Hl ,\! \Hhaton\!\! : & ~~s_n\! [m] \! = \! \widehat{h}_n x_n\! [m] \! + \! \underbracket{ \widetilde{h}_n  x_n\! [m] \! + \! {q[m]}  p[m] \! + \! v_n\! [m] }_{\text{new noise} ~\eta_{n,1}[m]},
\end{aligned}
\end{equation}
%
%
where the new noise terms depend on $\widetilde{h}_n$. Given $\ghatn$ at the AP, we obtain an achievable rate expression for a time slot by considering symbol-wise mutual information between channel input and output over the duration of  $\Nd$ data symbols as follows
\vspace{-1mm}
\begin{align} \label{RFormula}
R_n & \! = \! {W \Dd \over { \Nd}} \!\! \sum_{m=1}^{{ \Nd}} \!\! \bigg [ \betaon \,\mathbb{E} \left \{ I \!\left ( x_n[m]; s_n[m] \big | \ghatn, \Ho ,\Hhaton \right ) \right \} \nonumber  \\
&~~~~~~~~~~~~ \,+ \betaln \,\mathbb{E} \left \{ I \!\left (x_n[m]; s_n[m] \big | \ghatn, \Hl, \Hhaton \right ) \right \} \bigg ],
\end{align}
%
%
\noindent where $\Dd = \Td/\Tf$ is the fraction of the time slot used for data transmission and the expectations in \eqref{RFormula} are taken over the conditional probability density functions (pdfs) of $\ghatn$ given $\{\Heps, \Hhaton \}$ for $\varepsilon=0, 1$. To characterize $R_n$ in \eqref{RFormula} we need to find $\mathbb{E} \{ I (x_n[m]; s_n[m] \big | \ghatn, \Heps , \Hhaton) \}$. Exploiting the chain rule we can rewrite this expectation as follows

\vspace{-2mm}
\begin{align} \label{ECGivenH}
\mathbb{E} & \left \{I \! \left ( x_n[m]; s_n[m] \big | \ghatn, \Heps , \Hhaton \right ) \right \}  \\
& ~~~~~ = \sum_{k=0}^{K} \zeta_{k,n} \,I \! \left ( x_n[m]; s_n[m] \big | \ghatn, k, \Heps, \Hhaton \right ). \nonumber
\end{align}
%
Note that $I \big ( x_n[m]; s_n[m] \big | \ghatn, \Heps , \Hhaton \big )$ in \eqref{ECGivenH} is the mutual information between $x_n[m]$ and $s_n[m]$ when the battery state is $k$, given $\ghatn$ and $\{\Heps, \Hhaton \}$. From now on, we drop the variable $m$ in $x_n[m]$ and $s_n[m]$ for brevity of the presentation. Focusing on $I \big ( x_n; s_n \big | \ghatn, \Heps , \Hhaton \big )$, we  have
\vspace{-2mm}
\begin{align} \label{DEntropy}
I \!\left (x_n; s_n \big |\ghatn, k, \Heps, \Hhaton \right ) = &\,h \big ( x_n \big | \ghatn, k, \Hhaton, \Heps \big ) \nonumber \\
- &\,h \big ( x_n \big | s_n, \ghatn, k, \Hhaton, \Heps \big ),
\end{align}
%
where $h(\cdot)$ is the differential entropy.  Consider the first term in \eqref{DEntropy}. Since $x_n \sim {\cal CN} (0,P_{k,n})$ we have $ h \big ( x_n| \ghatn, k, \Hhaton,  \Heps \big ) = \log_2( \pi e P_{k,n})$. Consider the second term in \eqref{DEntropy}. Due to channel estimation error,  the new noises $\eta_{n,\varepsilon}$'s in \eqref{etaEquation} are non-Gaussian  and this term does not have a closed form expression. Hence, similar to \cite{Hassibi1,  Azadeh2, Vosoughi_Capacity}  we employ bounding techniques to find an upper bound on this term. This term is upper bounded by the entropy of a Gaussian random variable with the variance $\Theta_{\rm {M}}^{n, \varepsilon}$ 
\vspace{-2mm}
\begin{equation}
\Theta_{{\rm M}_{ } }^{n, \varepsilon} = \mathbb{E} \left \{ \big | x_n -\mathbb{E} \big \{ x_n \,| \,\ghatn, k, \Hhaton, \Heps \big \}  \big |^2 \right \},
\vspace{-1mm}
\end{equation}
%
%
where the expectations are taken over the conditional pdf of $x_n$ given $s_n,\ghatn, k, \Hhaton, \Heps$.  In fact, $\Theta_{\rm {M}}^{i, \varepsilon}$ is the mean square error (MSE) of the MMSE estimate of $x_n$ given $s_n,\ghatn, k, \Hhaton, \Heps$. Using minimum variance property of MMSE estimator, we have  $\Theta_{\rm {M}}^{n, \varepsilon} \leq \Theta_{\rm {L}}^{n, \varepsilon}$, where $\Theta_{\rm {L}}^{n, \varepsilon}$ is the MSE of the LMMSE estimate of $x_n$ given $s_n,\ghatn, k, \Hhaton, \Heps$.  Combining all, we find $h(x_n |s_n, \ghatn, k, \Hhaton, \Heps) \leq \log_2(\pi e \Theta_{\rm {L}}^{n, \varepsilon})$ and $I(x_n;s_n | \ghatn, k, \Hhaton,\Heps) \geq \log_2 (P_{k,n}/\Theta_{\rm {L}}^{n, \varepsilon})$ where
\vspace{-2mm}
\begin{align}
\Theta_{\rm {L}}^{n, \varepsilon} = &  \,\frac{P_{k,n} \sigma_{\eta_{n,\varepsilon}}^2 }{ \sigma_{ \eta_{n, \varepsilon}} ^2 + \ghatn P_{k,n}}, \\
 \sigma_{\eta_{n,\varepsilon}}^2 = &  \,{\gamtiln^ \varepsilon} P_{k,n} + \Sigmavn^2 + \varepsilon \, \Sigmap^2.
\end{align}
%
At the end, we obtain the lower bounds as follow
%
%
%
%
\vspace{-1mm}
\begin{subequations}\label{IH0H1}
	\begin{equation}
	\!\!\! I \!\left (\! x_n; s_n \big | \ghatn, k, \Hhaton, \Ho  \!\right ) \! \geq \log_2 \! \big(\!  1 \! + \ghatn b_{k,n}^0 \big),
	\vspace{-1mm}
	\end{equation}
	\begin{equation}
	\!\!\! I \!\left (\! x_n; s_n \big | \ghatn, k, \Hhaton, \Hl \! \right ) \! \geq \log_2\! \big(\! 1 \! + \ghatn b_{k,n}^1 \big),
	\vspace{-1mm}
	\end{equation}
\end{subequations}
%
where
\vspace{-2mm}
\begin{equation} 
b_{k,n}^0 = { P_{k,n} \over (\gamtiln^{0} P_{k,n} \! + \! \Sigmavn^2) }, ~~~	b_{k,n}^1 = { P_{k,n} \over (\gamtiln^{1} P_{k,n} \! + \! \Sigmavn^2 \! + \! \sigma^2_p) }.
\end{equation}
%
%
Substituting equations \eqref{ECGivenH} and \eqref{IH0H1} in \eqref{RFormula} and noting that the symbol-wise mutual information between channel input and output for $\Nd$ data symbols are equal we reach at

%
%
%
\vspace{-2mm}
\begin{align}\label{C_Ergodic}
R_n \! \geq \! R_{n, {\rm LB}}  \! = & \,{ \Dd} \betaon W \sum_{k=0}^{K} \zeta_{k,n} \mathbb{E} \Big \{ \! \log_2 \!\big(1+ \ghatn b^0_{k,n} \big ) | \Ho \Big \} \nonumber \\
+  & \,{ \Dd} \betaln W \sum_{k=0}^{K} \zeta_{k,n} \mathbb{E} \Big \{ \! \log_2 \!\big (1+ \ghatn b^1_{k,n} \big) | \Hl \Big \}.
\end{align}
%
%
Next, we compute the conditional expectations in \eqref{C_Ergodic}, in which we  take average over $\ghatn$, given $\Heps$. Using \eqref{Prob_alpha} and \eqref{a_c} we have
%
\begin{subequations}
\begin{align} \label{E_log}
\mathbb{E} \Big \{ \! \log_2 \!\big(1+ &\ghatn b^{\varepsilon }_{k,n} \big ) | \Heps \Big \} \nonumber \\
 &= \!\!\! \sum_{i=1}^{\lfloor k \Omega_n \! \rfloor \! - \! \alphatra} \!\!\! \int_{a_{i,k,n}} ^{c_{i,k,n}} \!\! \log_2 \! \big(1 + \text{S}^{\varepsilon }_{i,n} \,x \big ) f^{\varepsilon }_{\ghatn} \!(x) dx \nonumber \\
 &= \!\!\! \sum_{i=1}^{\lfloor k \Omega_n \! \rfloor \! - \! \alphatra} \! V_k(\text{S}^{\varepsilon }_{i,n} ,\gamhatn^ \varepsilon)
\end{align}
%
in which 
%
\hspace{-4mm}
\begin{align} 
\text{S}_{i,n}^0 \! = \! { i \,\pu \over (\gamtiln^{0} \,i \,\pu \! + \! \Sigmavn^2) }, ~~~~	\text{S}_{i,n}^1 \! = \! { i \,\pu \over (\gamtiln^{1} \,i \,\pu \! + \! \Sigmavn^2 \! + \! \sigma^2_p) },
\end{align}
\begin{align} 
V_k(\text{S}_{i,n},\gamhatn) \! = \! M(c_{i,k,n}, \text{S}_{i,n},\gamhatn ) \! - \! M(a_{i,k,n}, \text{S}_{i,n} ,\gamhatn),
\end{align}
\end{subequations}
%
and
%
\begin{align} 
M(x,\text{S}, &w) = \!\! \int \! \log_2(1 + \text{S} x) { e^{-x \over w} \over w } dx \nonumber \\
= & \, \frac {e^{1 \over \text{S} w} } {\ln(2)} ~\text{Ei} \Big ( {-x \over w} - { 1 \over \text{S} w } \Big ) - e^{-x \over w}\log_2(1+\text{S} x) .
\end{align}
%
%
Also, $c_{i,k,n}$ and $a_{i,k,n}$ are given in \eqref{a_c}. Substituting \eqref{E_log} in \eqref{C_Ergodic} we reach to
%
%
\begin{align} 
R_{n,{\rm LB}} \! = & \,\Dd \beta_{0, n} W \! \sum_{k= \alphatra +1}^K \!\!\!\!\! \sum_{i= 1} ^{\lfloor k \Omega_n\! \rfloor \! - \! \alphatra} \!\! \zeta_{k,n} V_k( \text{S}_{i,n}^0, \gamhatn^0) \nonumber \\ 
+ & \,\Dd \beta_{1, n} W \! \sum_{k= \alphatra +1}^K \!\!\!\!\! \sum_{i= 1} ^{\lfloor k \Omega_n\! \rfloor \! - \! \alphatra} \!\! \zeta_{k,n} V_k( \text{S}_{i,n}^1, \gamhatn^1).
\end{align}
%
\noindent We note that the lower bounds in \eqref{IH0H1} are achieved when the new noises $\eta_{n,0},\eta_{n,1}$ in \eqref{etaEquation} are regarded as worst-case Gaussian noise and hence the MMSE and LMMSE of $x_n$ given $s_n, \ghatn, k, \Hhaton, \Heps$ coincide. Given the rate lower bound $R_{n, {\rm LB}}$ for SU$_n$, the sum-rate lower bound for all SU$_n$'s is 
%
\begin{equation}\label{AvgSumRate}
\RLB = \sum_{n=1}^{\Nu} R_{n,{\rm LB}}.
\end{equation}
%
%
\par So far, we have established a sum-rate lower bound on the achievable sum-rate. Next, we characterize the average inference constraint (AIC). Suppose $\Ibar$ is the maximum allowed average interference power, i.e.,  the average interference power that collective SUs impose on \PUrx ~cannot exceed $\Ibar$. To satisfy AIC, we have
%
\begin{equation}\label{AIC0}
\sum_{n=1}^{\Nu} \betaln \mathbb{E} \{ z_n \} \Big [ \Dd  \,\mathbb{E} \big \{ P_n( \ghatn )\big \} + \Dt \Ptra \Big ] \leq \Ibar,
\end{equation}
%
where $\Dt = \Ttra/\Tf$ and the expectation is over the conditional pdfs of $\ghatn$ under $ \{ \Hl, \Hhaton\}$. The first term in \eqref{AIC0} is the average interference power imposed on \PUrx ~when SUs transmit data symbols, and the second term  is the average interference imposed on \PUrx ~when SUs send training symbols for channel estimation at the AP. Using \eqref{Prob_alpha} we compute the term with expectation inside \eqref{AIC0} as follows
%
\begin{align} \label{AIC1}
\mathbb{E} \big \{ P_n( \ghatn ) \big \} = &\sum_{k =0}^{K} \zeta_{k,n} \sum_{i=0}^{K} \Pr(\alpha_{k,n} = i | \Hl ) \,i \,\pu  \nonumber\\
= & \sum_{k=\alphatra \! + \! 1}^{K} \zeta_{k,n} \!\!\! \sum_{i=1}^{ \lfloor k \Omega_n \!\rfloor \! - \! \alphatra} \!\! \psi_{i,k,n}^1 \,i \,\pu.
\end{align}
%
Substituting \eqref{AIC1} into \eqref{AIC0}, we can rewrite the AIC in \eqref{AIC0} as
%
\begin{equation} \label{AIC}
\sum_{n=1}^{\Nu} \! \betaln \delta_{z_n} \! \bigg [ \! \sum_{k=\alphatra \! + \! 1}^{K} \!\!\! \zeta_{k,n} \!\!\!\!\! \sum_{i=1}^{ \lfloor k \Omega_n \!\rfloor \! - \! \alphatra} \! \!\!\psi^1_{i,k,n} \,i \,\pu + \Dt \Ptra \bigg ] \leq \Ibar.
\end{equation}
%
\noindent For ideal spectrum sensing we get $\betaln=0$ in \eqref{Beta}, implying that  data transmission from SUs to the AP does not cause interference on \PUrx ~and the left-hand side of \eqref{AIC} becomes zero, i.e., the AIC is always satisfied. 
\par Next, we  examine how spectrum sensing error and channel estimation error affect $\RLB$ and AIC expressions. {First}, spectrum sensing error affects AIC via $\betaln$, and $\RLB$ via $\betaon$ and $\betaln$. Recall $\betaon,\betaln$ depend on $\pi_0, {P}_{{\rm fa}_n}, {P}_{{\rm d}_n} $ (see \eqref{Beta}).  {Second}, channel estimation error affects AIC via {$\Dt, \psi_{i,k,n}^1$}, and  $\RLB$ via $\gamtiln^\varepsilon$. 

\par Having the mathematical expressions for $\RLB$ and AIC, our goal is to optimize the set of transmission parameters $\{\Omega_{n}, \theta_{n}\}$ for all SUs such that $\RLB$ is maximized, subject to the AIC. To inspect the underlying trade-offs between decreasing the average interference power imposed by SU$_n$'s on \PUrx ~and increasing the sum-rate lower bound $\RLB$, we note that increasing data symbol transmission power $P_{k,n}$ increases $\RLB$. However, it increases the average interference power. Aiming to strike a balance between increasing $\RLB$ and decreasing the imposed average interference power, we seek the optimal $\{\Omega_{n}, \theta_{n} \}_{n=1}^{ \Nu}$ such that $\RLB$ in \eqref{C_Ergodic} is maximized, subject to AIC given in \eqref{AIC}. In other words, we are interested in solving the following constrained optimization problem 
\leqnomode
\begin{align}\tag{P1}\label{Prob1}
& \!\!\!\! {\underset { \{\Omega_{n} , ~\theta_{n} \}_{n=1}^{ \Nu} }  {\text{Maximize}} }  ~\RLB \\
\text{s.t.:}  ~~& \Omega_{n} \in [0,1], \nonumber \\
& \theta_{n} \geq 0 , \nonumber \\
& \text{AIC in} ~\eqref{AIC}  ~\text{is satisfied.} \nonumber
\end{align}
\reqnomode
We note that solving \eqref{Prob1} requires an $2 \Nu$-dimensional search over the search space $[0,1]^{\Nu}  \times {[0, \infty)}^{ \Nu}$.
%
%
%
\section{Simulation Results}\label{SimResults}

\par In this section we corroborate our analysis on constrained maximization of the achievable sum-rate lower bound with Matlab simulations, and examine how the optimized sum-rate lower bound depends on the average number of harvesting energy packets $\rho_n$, the maximum allowed average interference power $\Ibar$, the duration of spectrum sensing phase $\Tsen$, the number of consumed cells of energy for channel probing $\alphatra$, and the size of the battery $K$. Our simulation parameters are given in Table \ref{table2}.

$\bullet$  \textbf {Spectrum sensing-channel probing-data transmission trade-offs}: To explore these trade-offs, in this section we let $\Nu=1$ and examine how the rate lower bound {$\RLB$} in \eqref{AvgSumRate} for a single user changes when we vary $\Tsen$, or $\alphatra$. The simulation parameters, except for $\alphatra, \Tsen, \sigma_w^2, \sigma_v^2$ are given in Table \ref{table2}
\footnote{Note that the variances of channel estimate and corresponding estimation error in \eqref{var_H} depend on the product $\Ptra \Ntra = \alphatra \eu \fs$ and is independent of $\Ttra$. That is the reason, instead of $\Ttra$, we consider varying $\alphatra$, to understand channel probing trade-offs.}.

\par Fig. \ref{C_vs_Tsen} shows {$\RLB$} versus $\Tsen$ for two values of the energy harvesting parameter $\rho= 15, 16, \sigma_w^2 \! = \! \sigma_v^2 \! = \! 1$ and $\alphatra=1$. This figure suggests that there exists a trade-off between $\Tsen$ and {$\RLB$}. On the positive side, as $\Tsen$ (or equivalently $\Nsen$) increases, the accuracy of the energy detector for spectrum sensing increases (i.e., $P_{{\rm fa}_n}$ in \eqref{P_fn} decreases). 
A more accurate spectrum sensing can reveal new opportunities for SU$_n$ to be exploited for its data transmission, that can increase {$\RLB$}. On the negative side, as $\Tsen$ increases, the duration of data transmission phase $\Td= \Tf - \Tsen - \Ttra$ decreases. This trade-off between spectrum sensing and data transmission indicates that, given the parameters (including $\alphatra$), there is an optimal $\Tsen$, denoted as $\Tsen^*$ in Fig. \ref{C_vs_Tsen},  that maximizes {$\RLB$}. For instance, for $\rho=15,16$ we have $\Tsen^*= 0.6, 0.75$\,ms.

\par Fig. \ref{C_vs_a_Tra} plots {$\RLB$} versus $\alphatra$ for $\rho= 18, 20$, $\Tsen=1$\,ms and $\sigma_w^2 \! = \! \sigma_v^2 \! = \! 5$. This figure suggests that a trade-off exists between $\alphatra$ and $\RLB$. On the positive side, as $\alphatra$  increases, the accuracy of channel probing (measured by the variance of channel estimation error in \eqref{var_H}) improves. A more accurate channel probing can increase $\RLB$. On the negative side, as $\alphatra$  increases, the available energy for data transmission decreases. This trade-off between channel probing and data transmission shows that, given the parameters (including $\Tsen$), there is an optimal $\alphatra$, denoted as $\alphatra^*$ in Fig. \ref{C_vs_a_Tra},  that maximizes {$\RLB$}. For instance, for $\rho=15,16$ we have $\alphatra^*=4$. 
%
%
\begin{table}[t!]
	\begin{center}
		\vspace{0mm}
		\caption{Simulation Parameters}
		\label{table2}
		\begin{footnotesize}
			\begin{tabular}{|l|l||l|l|} 
				\hline
				\textbf{Parameter} & \textbf{Value} & \textbf{Parameter} & \textbf{Value} \\
				\hline \hline
				$\Pp$ & $1\,$watts & $\Sigmavn^2$ & $1$  \\
				$\pi_0$ & $0.7$ & $\Sigmawn^2$ & $1$   \\
				$\Ttra$ & $0.1\,$ms &  $\alphatra$ & $1$   \\
				$\Tsen$ & $1\,$ms & ${\overline{P}}_{\rm{d}}$ & $0.85$  \\
				$\Tf$ & $10\,$ms &  $W$ & $10\,$KHz  \\
				$\eu$ & $0.01$ & $\delta_{q}$ & $1$  \\
				\hline
			\end{tabular}
		\end{footnotesize}   
	\end{center}    
	\vspace{0mm}
\end{table}
%
%
%
\begin{figure}[!t] 
	\centering
	\hspace{-0mm}
	\begin{subfigure}[b]{0.5\textwidth}                
		\centering	         
		\includegraphics[width=60mm]{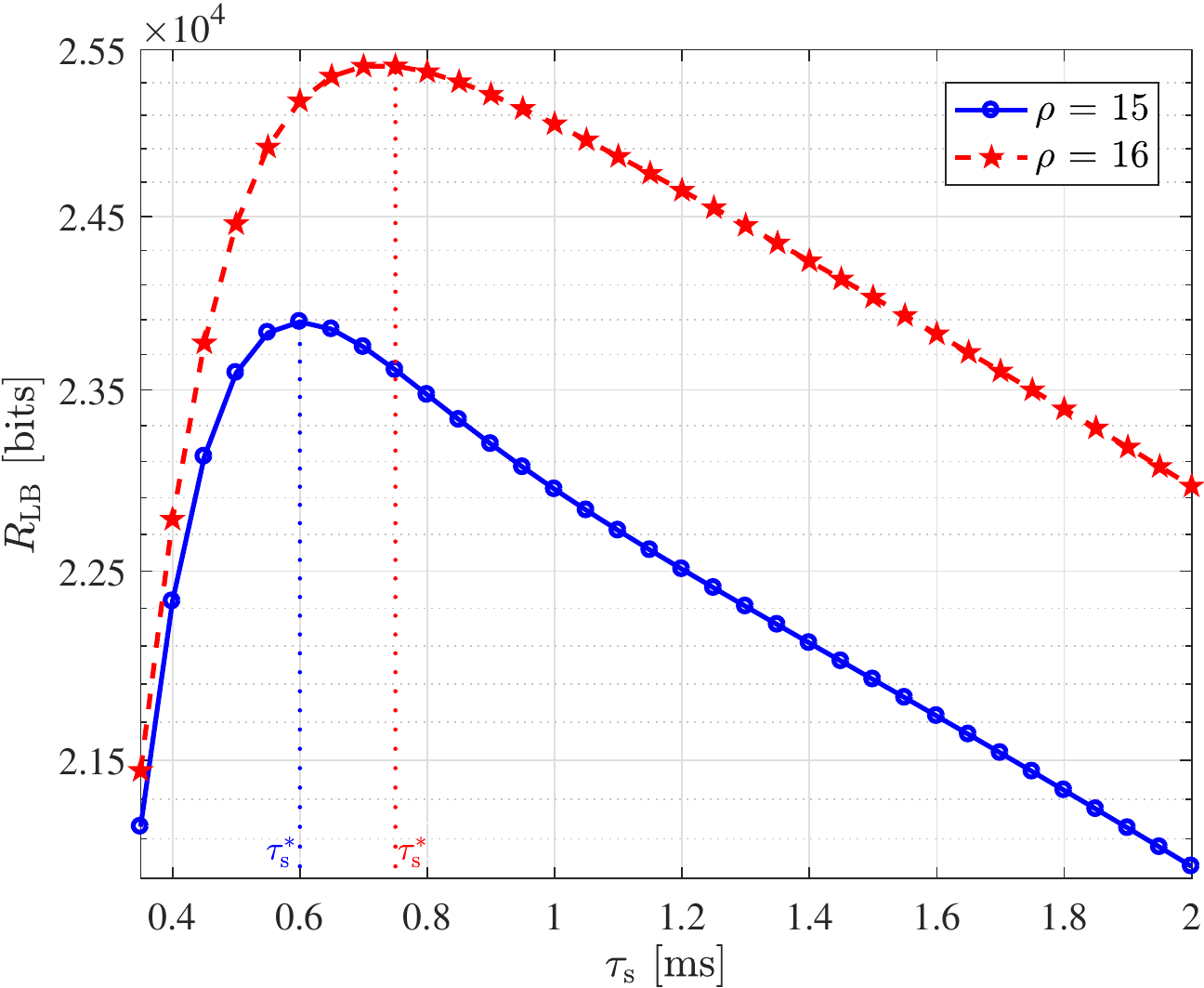}
		\caption{} 
		\label{C_vs_Tsen}   
		\vspace{2mm}                
	\end{subfigure} 
	\hspace{-0mm}
	\begin{subfigure}[b]{0.5\textwidth}
		\centering      
		\includegraphics[width=60mm]{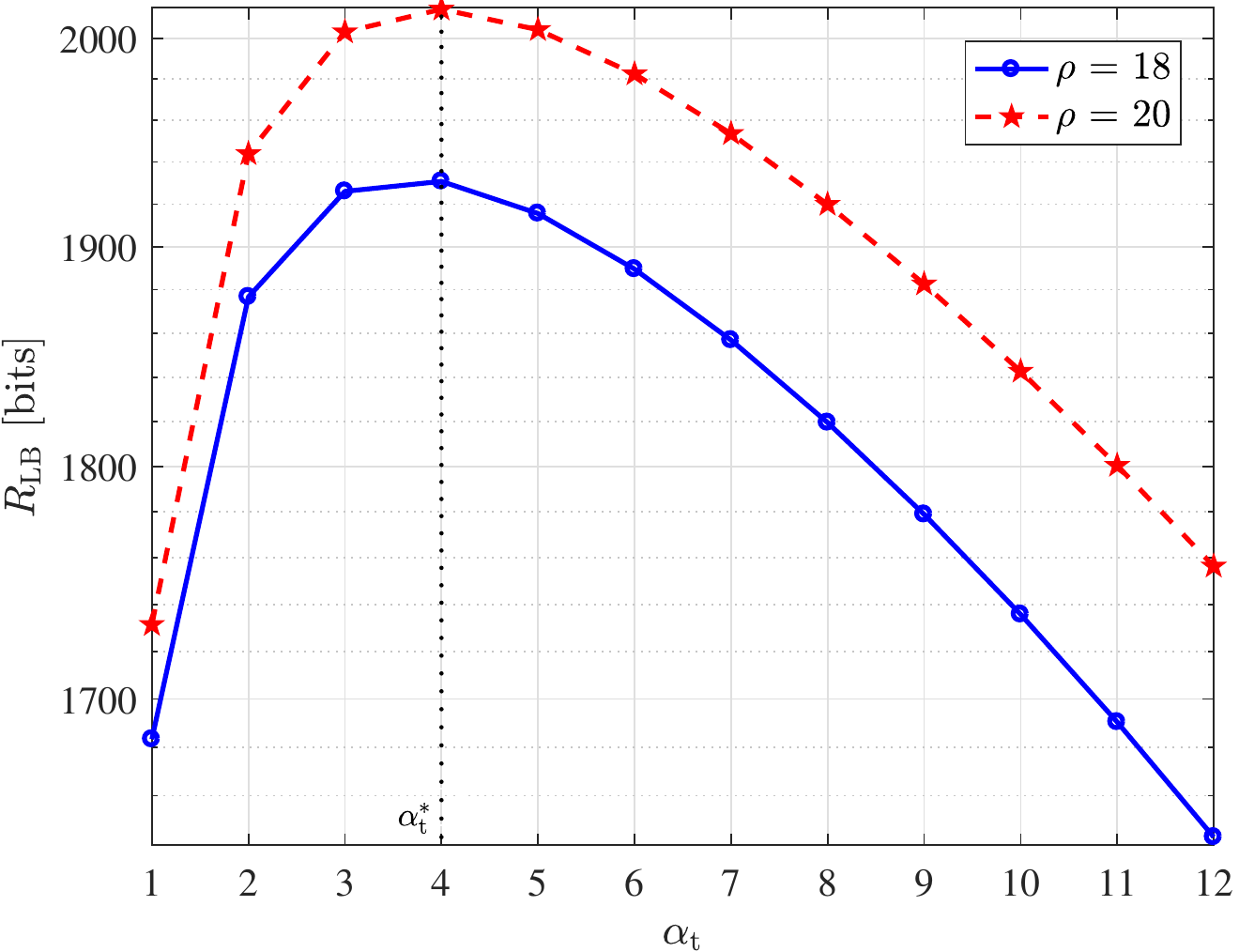}
		\caption{} 
		\label{C_vs_a_Tra}    
	\end{subfigure} \\
	\caption{(a) $\RLB$ versus $\Tsen$ for $K=80, \theta=0.25, \Omega=0.35, \sigma_w^2 \! = \! \sigma_v^2 \! = \! 1,$    (b) $\RLB$ versus $\alphatra$ for $K=200, \theta=0.25, \Omega=0.35, \sigma_w^2 \! = \! \sigma_v^2 \! = \! 5.$}
	\label{C_vs_T}
\end{figure}
%
%
%
%
%
%
%
\par $\bullet$ \textbf {Effect of the optimization variables $\boldsymbol{\Omega, \theta}$}: In this section, we let $\Nu=1$ and we illustrate how the entries of the steady-state probability vector $\boldsymbol{\zeta}$ in \eqref{zeta_fin}, $\RLB$ in \eqref{AvgSumRate} for a single user, and the battery outage probability  $P_{b}^{\rm{Out} }$ defined below depend on the optimization variables $\Omega$ and $\theta$. We define $P_{b_n}^{\rm{Out} }$ as the steady-state probability of the battery of SU$_n$ being equal or lower than $\alphatra$. When the battery is at outage, it cannot yield energy for data transmission or channel probing. We have
%
\begin{equation} \label{Pr_battery_out}
P_{b_n}^{\rm{Out} } =\Pr( {\cal B}_n \leq \alphatra) = \sum_{k=0}^{\alphatra} \zeta_{k,n}.
\end{equation}
%
%
The simulation parameters are given in Table \ref{table2}. Also, we let  $\boldsymbol{ \gamma } = 2, \boldsymbol{ \delta }_u = 1 , \boldsymbol{ \delta }_z = 1$. 

\par Fig. \ref{C_vs_Omega} illustrates $\RLB$ for a single user versus $\Omega$ for $\rho=15, 20$. We observe that $\RLB$ is neither  a convex nor a concave function of $\Omega$. This figure suggests that there is an optimal $\Omega$, which we denote as $\Omega^*$, that maximizes $\RLB$.  Starting from small values of $\Omega$, as $\Omega$ increases (until it reaches the value $\Omega ^*$), $\RLB$ increases, because the harvested energy can recharge the battery and can yield more power for data transmission. However, when $\Omega$ exceeds $\Omega ^*$, the harvested and stored energy cannot support the data transmission and $\RLB$ decreases. Moreover, as $\rho$ increases, $\RLB$ increases as well. 
The behavior of $\RLB$ versus $\theta$ is shown in Fig. \ref{C_v_theta_fig} for $\rho= 15, 18$. We observe that $\RLB$ is neither a convex nor a concave function of $\theta$. Similar to $\Omega$, there is an optimal $\theta$, which we denote as $\theta^*$, that maximizes $\RLB$.  Starting from small values of $\theta$, as $\theta$ increases (until it reaches $\theta^*$), $\RLB$ increases. However, when $\theta$ exceeds $\theta^*$, $\RLB$ decreases.
%
%
\begin{figure}[!t] 
	\centering
	\hspace{-0mm}
	\begin{subfigure}[b]{0.5\textwidth}                
		\centering	         
		\includegraphics[width=60mm]{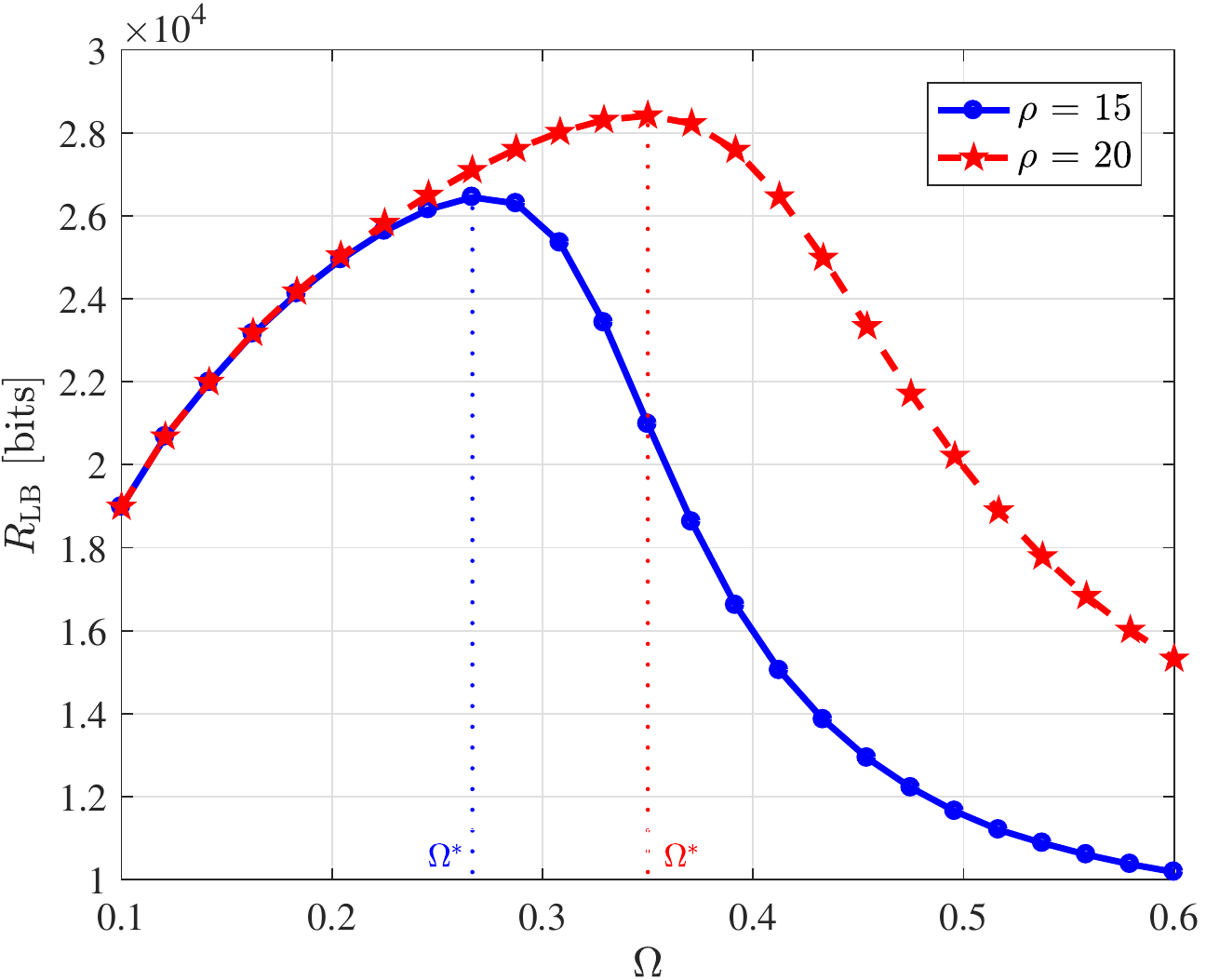}
		\caption{} 
		\label{C_vs_Omega}   
		\vspace{2mm}                
	\end{subfigure} 
	\hspace{-0mm}
	\begin{subfigure}[b]{0.5\textwidth}
		\centering      
		\includegraphics[width=60mm]{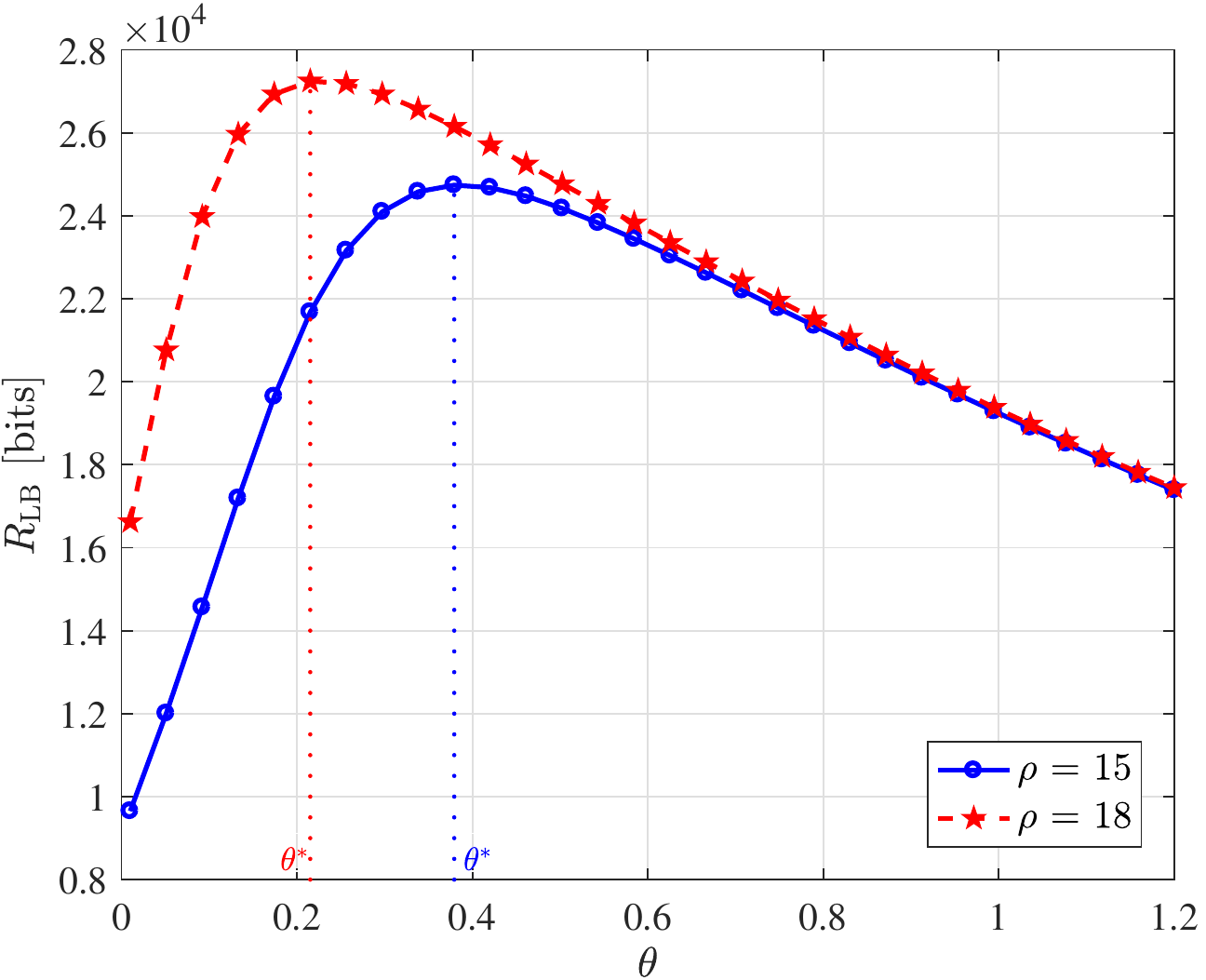}
		\caption{} 
		\label{C_v_theta_fig}    
	\end{subfigure} \\
	\caption{(a) $\RLB$ versus $\Omega$ for $K=80, \theta=0.2,$  (b) $\RLB$ versus $\theta$ for $K=80, \Omega=0.35.$ }
	\label{Fig6ab}
\end{figure}
%
%
%
%
%
%
\par Fig. \ref{Pr_b6} plots the entries of the steady-state probability vector $\boldsymbol{\zeta}$ versus $k$ for $\Omega=0.45, 0.3$ and $\theta=0.2$. Fig. \ref{Pr_b8} plots the entries of $\boldsymbol{\zeta}$ versus $k$ for $\theta=0.1, 0.5$ and $\Omega=0.35$. To quantify the effect of $\Omega$ and $\theta$ on the entries of $\boldsymbol{\zeta}$ we define the average energy stored at the battery of SU$_n$ as
%
%
\begin{equation} 
\overline{ \cal B }_n = \mathbb{E} \{ {\cal B}_n \} = \sum_{k=0}^{K} k \,\zeta_{k,n},
\end{equation}
%
%
where the largest possible value for $\overline{ \cal B }_n$ is $K$. Considering Figs. \ref{Pr_b6_a} and \ref{Pr_b6_b}, we find $\overline{ \cal B }^{(a)} = 16.97$ for $\Omega=0.45$ (the battery is near empty) and $\overline{ \cal B }^{(b)} = 66.30$ for $\Omega=0.30$ (the battery is near full).  Considering Figs. \ref{Pr_b8_a} and \ref{Pr_b8_b}, we find $\overline{ \cal B }^{(a)} = 24.08$ for $\theta=0.1$ and $\overline{ \cal B }^{(b)} =  71.55$ for $\theta=0.5$.
Clearly, the values of $\Omega$ and $\theta$ affect $\overline{ \cal B }$. Given $\theta$, when $\Omega$ is large, data transmit  energy {$\alpha_{k}$} in \eqref{alphaBAR} is large. Due to large energy consumption for data transmission (compared to energy harvesting) the battery becomes near empty at its steady-state and SU may stop functioning, due to energy outage. When $\Omega$ is small, {$\alpha_{k}$} in \eqref{alphaBAR} is small.  Due to small energy consumption for data transmission (compared to energy harvesting) the battery becomes near full at its steady-state, indicating that SU has failed to utilize the excess energy. Both cases inevitably hinder data transmission, leading to a reduction in $\RLB$. Similar argument holds true, when $\theta$ varies and $\Omega$ is given. In particular, when $\theta$ is small, transmit energy {$\alpha_{k}$} in \eqref{alphaBAR} is large, and when $\theta$ is large, transmit energy {$\alpha_{k}$} in \eqref{alphaBAR} is small. Again, both cases impede data transmission, leading to a lower $\RLB$. Overall, the observations we make in Figs. \ref{Fig6ab}, \ref{Pr_b6}, \ref{Pr_b8} confirm that optimizing both $\Omega$ and $\theta$ to achieve a balance between the energy harvesting and the energy consumption for data transmission is of high importance.
%
%
%
%
\begin{figure}[!t] 
	\centering
	\hspace{-0mm}
	\begin{subfigure}[b]{0.5\textwidth}                
		\centering	         
		\includegraphics[width=60mm]{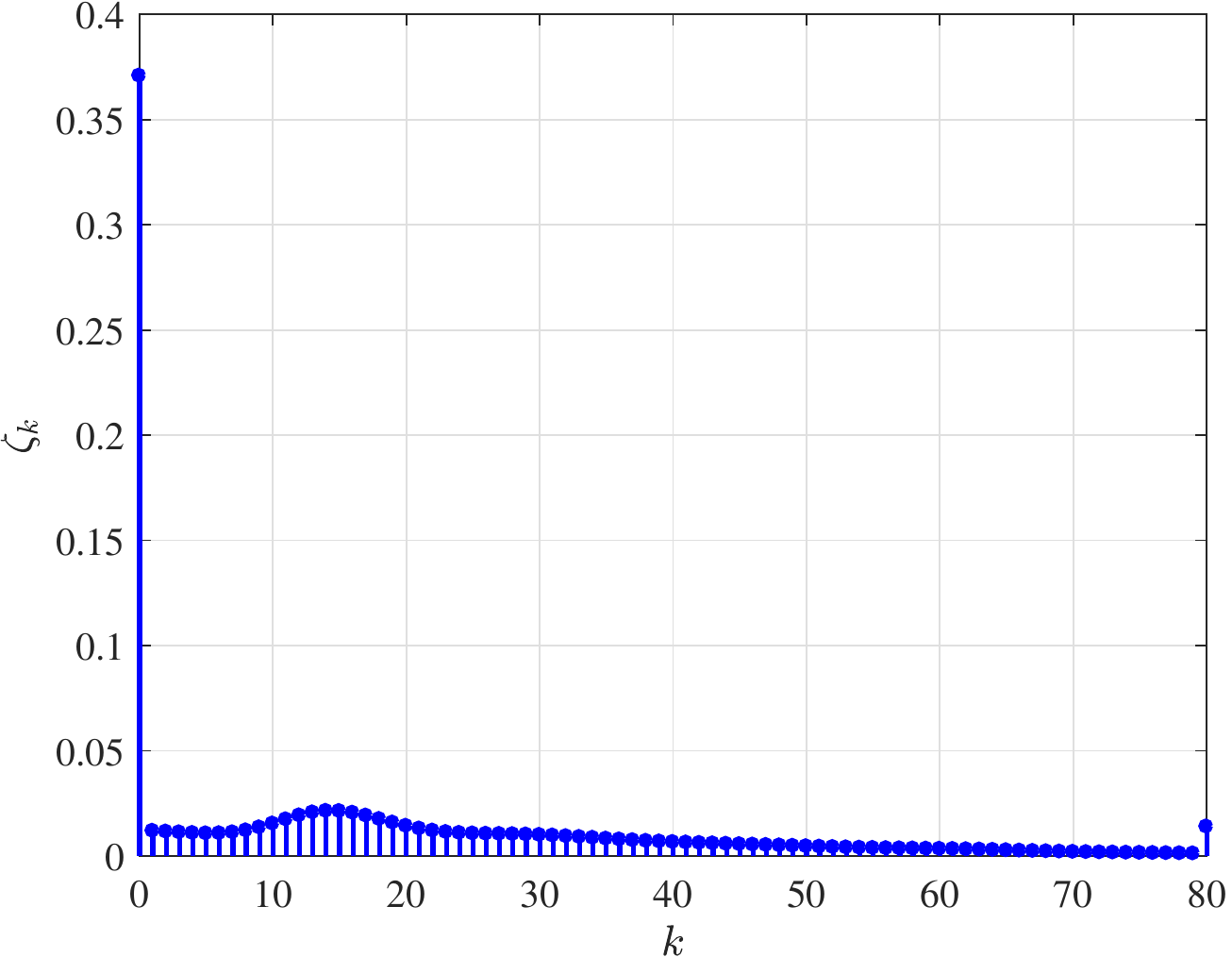}
		\caption{} 
		\label{Pr_b6_a}                   
	\end{subfigure} 
	\hspace{-0mm}
	\begin{subfigure}[b]{0.5\textwidth}
		\centering      
		\includegraphics[width=60mm]{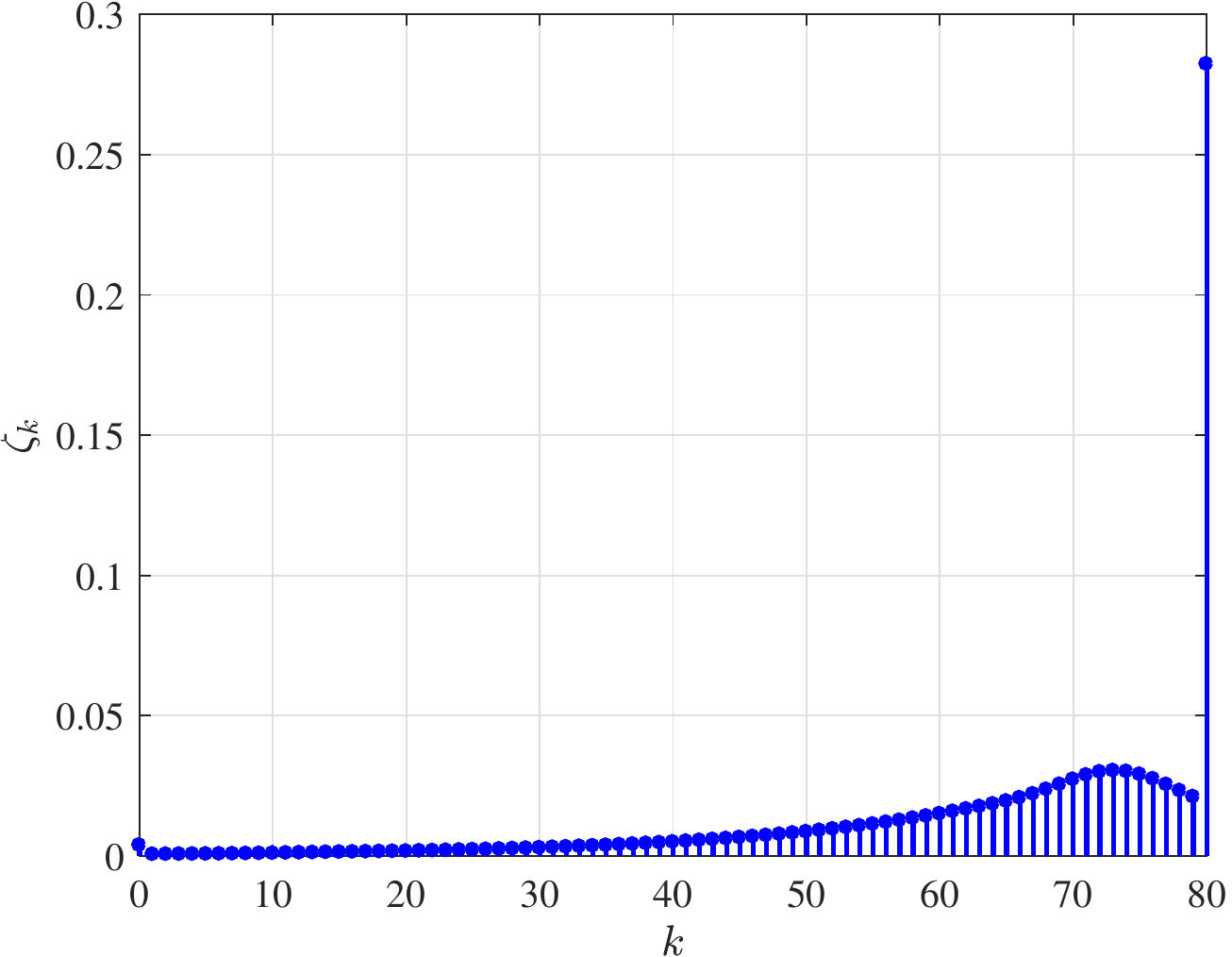}
		\caption{} 
		\label{Pr_b6_b}    
	\end{subfigure} \\
	\caption{$\zeta_{k}$ versus $k$ for $K=80, \rho= 15, \theta=0.2$, (a) $\Omega = 0.45$ , (b) $\Omega=0.30$.}
	\label{Pr_b6}
\end{figure}
%
%
%
%
\begin{figure}[!t] 
	\centering
	\begin{subfigure}[b]{0.5\textwidth}                
		\centering	         
		\includegraphics[width=60mm]{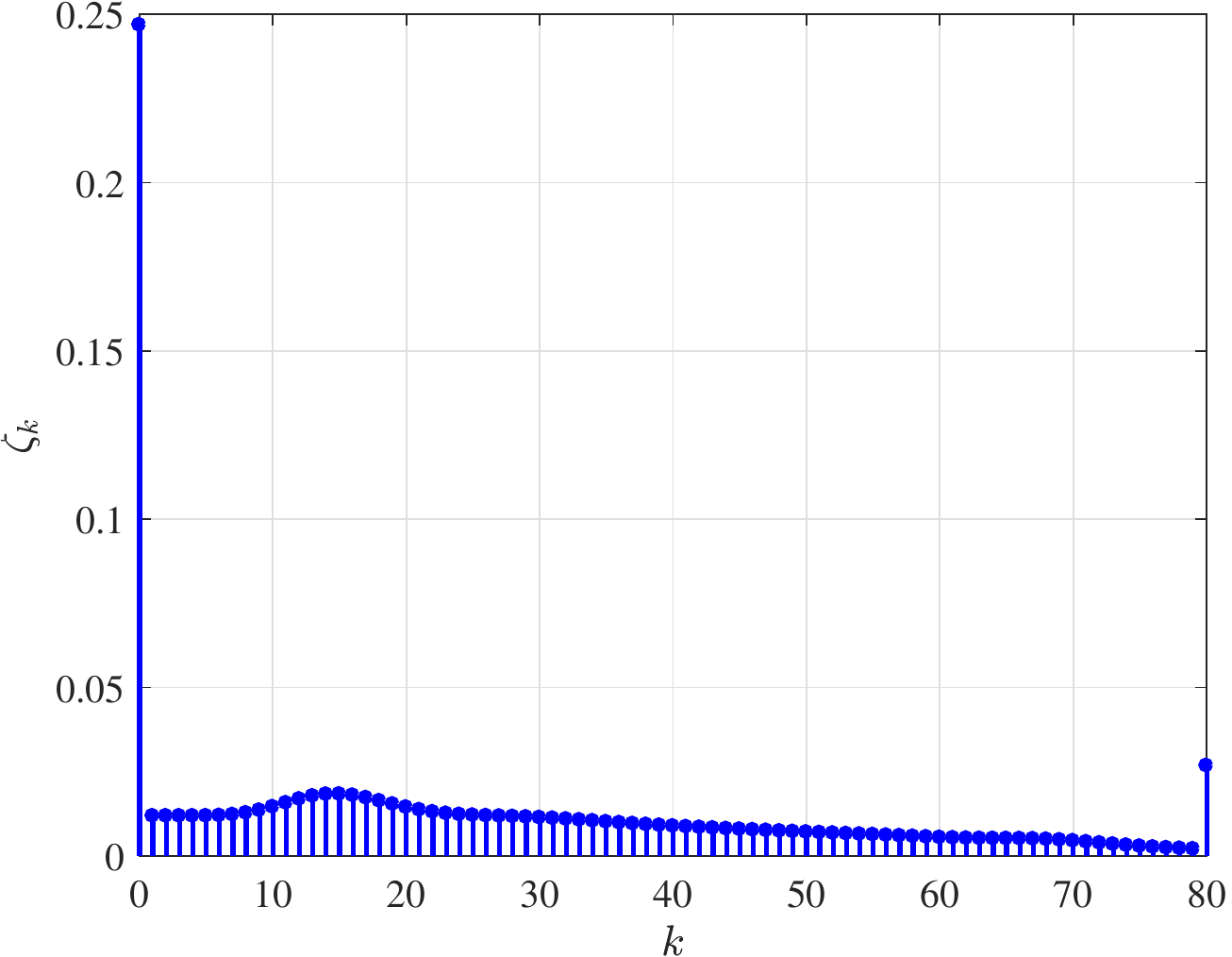}
		\caption{} 
		\label{Pr_b8_a}                   
	\end{subfigure} 	
	\hspace{-4mm}
	\begin{subfigure}[b]{0.5\textwidth}
		\centering      
		\includegraphics[width=60mm]{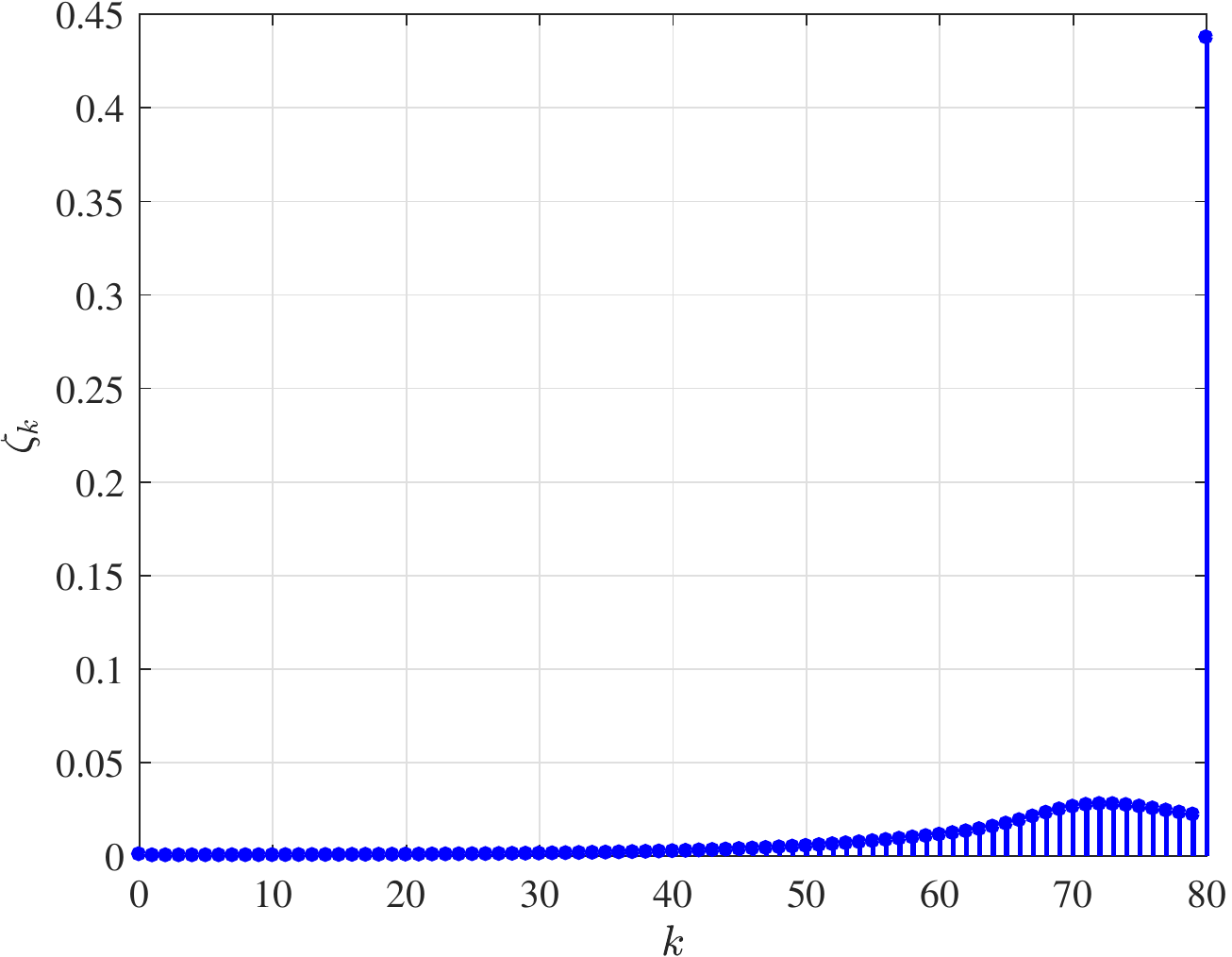}
		\caption{} 
		\label{Pr_b8_b}    
	\end{subfigure} \\	
	\caption{$\zeta_{k}$ versus $k$ for $K=80,  \rho= 15 ,\Omega=0.35$, (a) $\theta = 0.1$ , (b) $\theta=0.5$.}
	\label{Pr_b8}
\end{figure}
%
%
\par Fig. \ref{Pb_Out_vs_Omega} illustrates the behavior of $P_{b}^{\rm{Out} }$ for a single user in terms of $\Omega$ for $\theta=0.05$. Fig. \ref{Pb_Out_vs_theta} plots $P_{b}^{\rm{Out} }$ versus $\theta$ for $\Omega=0.35$. For $\alphatra=1$, $P_{b}^{\rm{Out} }$ in \eqref{Pr_battery_out} reduces to $P_{b}^{\rm{Out} } =\zeta_0+\zeta_1$, i.e.,  $P_{b}^{\rm{Out} }$ depends on $\Omega$ and $\theta$, via only the first two entries of vector $\boldsymbol{\zeta}$. 
Fig. \ref{Pb_Out_vs_Omega} shows that, as $\Omega$ increases, $P_{b}^{\rm{Out} }$ increases as well. This is because as $\Omega$ increases, given $\theta$, {$\alpha_k$} in  \eqref{alphaBAR} increases. Due to large energy consumption for data transmission the chance of the battery depletion and hence $P_{b}^{\rm{Out} }$ increase. 
Fig. \ref{Pb_Out_vs_theta} demonstrates that, as $\theta$ increases, $P_{b}^{\rm{Out} }$ decreases. This is because as $\theta$ increases, given $\Omega$, {$\alpha_k$} in \eqref{alphaBAR} decreases. Due to small energy consumption for data transmission the chance of the battery depletion and hence $P_{b}^{\rm{Out} }$ decrease.
%
%
%
\begin{figure}[!t] 
	\centering
	\hspace{-0mm}
	\begin{subfigure}[b]{0.5\textwidth}                
		\centering	         
		\includegraphics[width=60mm]{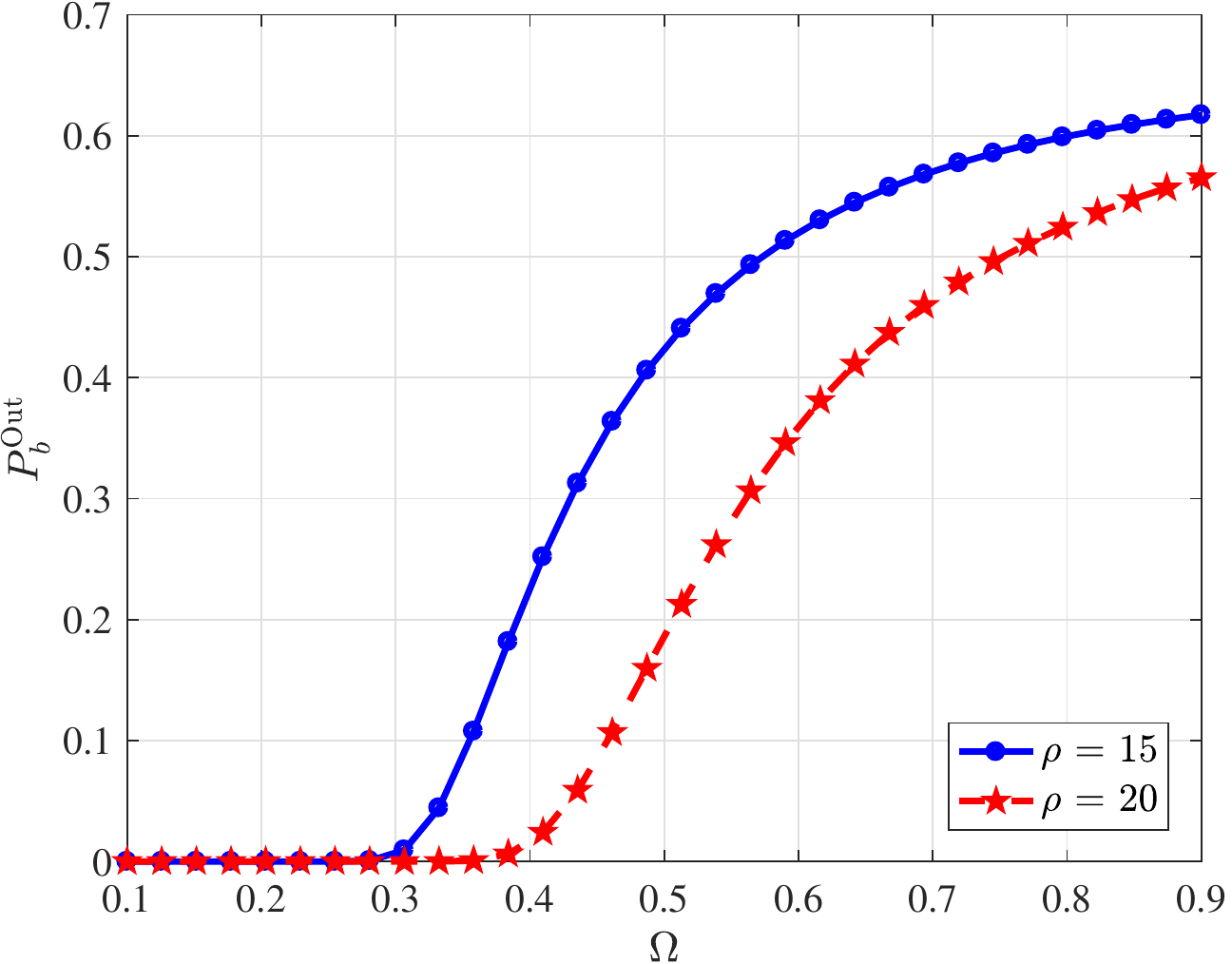}
		\caption{} 
		\label{Pb_Out_vs_Omega}   
		\vspace{2mm}                
	\end{subfigure} 
	\hspace{-0mm}
	\begin{subfigure}[b]{0.5\textwidth}
		\centering      
		\includegraphics[width=60mm]{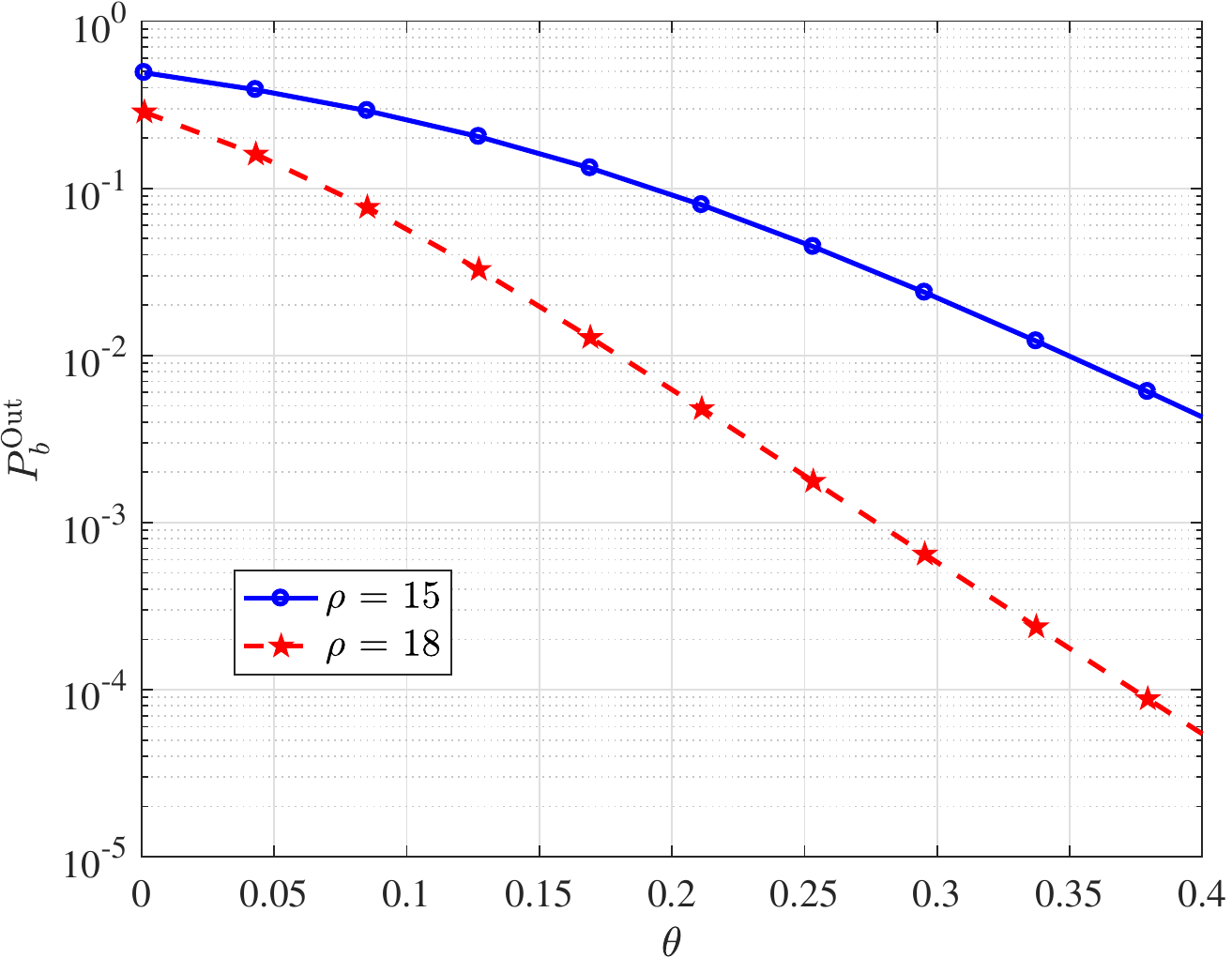}
		\caption{} 
		\label{Pb_Out_vs_theta}    
	\end{subfigure} \\
	\caption{(a) $P_{b}^{\rm{Out} }$ versus $\Omega$ for $K=80, \theta=0.05$,  (b) $P_{b}^{\rm{Out} }$ versus $\theta$ for $K=80, \Omega=0.35$.}
	\label{}
\end{figure}
%
%
%
%
%
%
%
\par $\bullet$ \textbf{Solving Problem (\ref{Prob1})}: Next, we consider solving the constrained optimization problem  \eqref{Prob1} and plot the maximized $\RLB$, denoted as $\RLB^*$ ($\RLB^*$ is $\RLB$ evaluated at the solutions obtained from solving  \eqref{Prob1}). 

\par Fig. \ref{C_vs_K_fig} depicts  $\RLB^*$ versus $K$ for $\Nu=3$. We let the statistics of fading coefficients be different across SUs, $\boldsymbol{ \gamma } = [2 , 2.2 , 2.1], \boldsymbol{ \delta }_u = [1 , 0.8 , 1.2], \boldsymbol{ \delta }_z = [1 , 0.5 , 0.8]$ and $\rho = 30, 40$ be equal for all SUs. We observe that as $K$ increases, $\RLB^*$ increases. This is expected, since as $K$ increases the chance of energy overflow decreases, leading to a larger amount of stored energy in the battery, which can be utilized to support a higher data rate transmission.
%
\begin{figure}[!t]
\centering       			
\includegraphics[width=60mm]{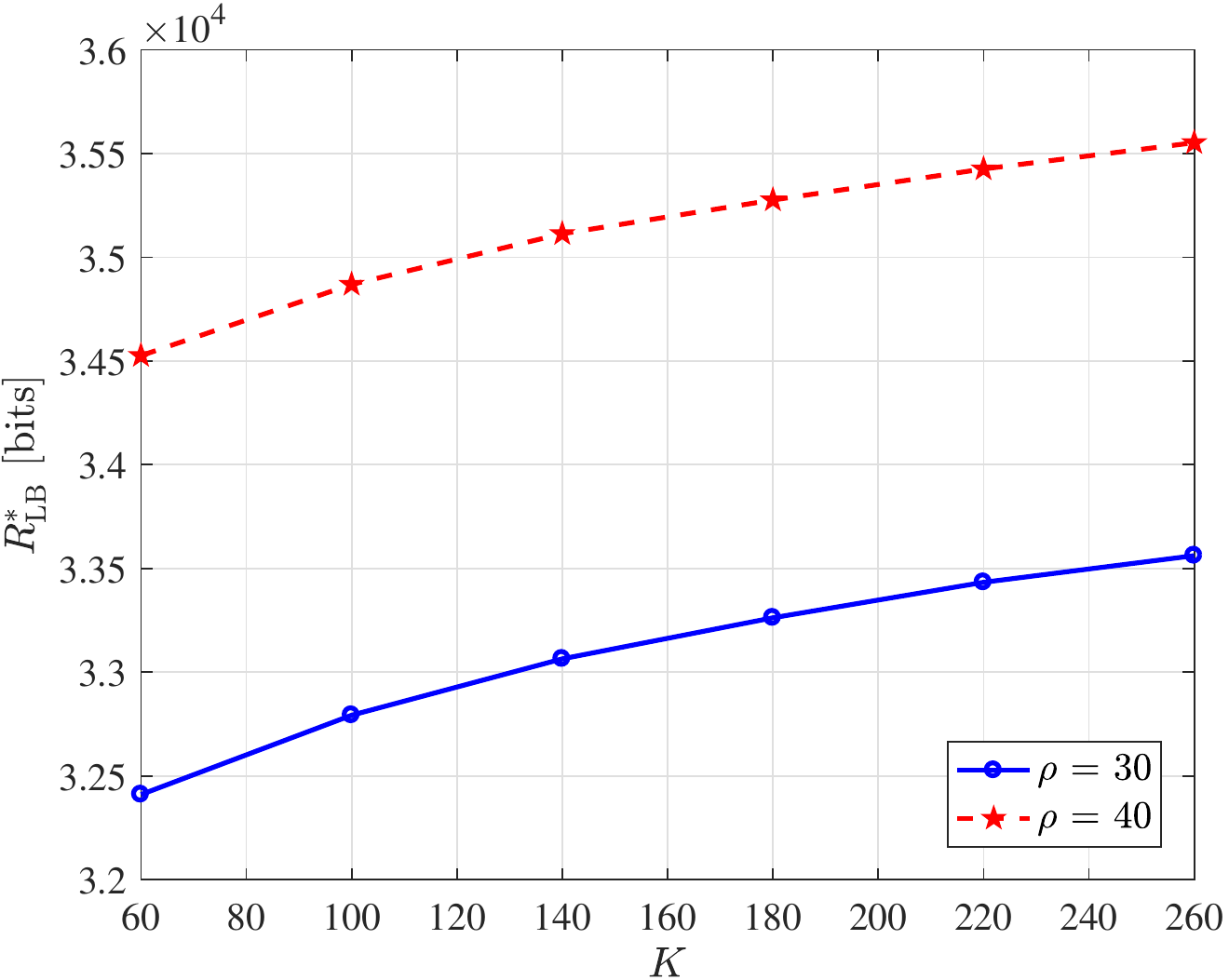}
\caption{$\RLB^*$ versus $K$ for $\Ibar=2$\,dB.} 
\label{C_vs_K_fig}      
\vspace{-0mm}
\end{figure}
%
%
%
\par Fig. \ref{C_3Nu_Iav2} shows $\RLB^*$ versus $\Ibar$ for $K=80$, $\rho=10, 15$ and {$\Nu=3$}. For small $\Ibar$, the AIC in \eqref{AIC} is active and consequently, it limits transmit power of SUs. As $\Ibar$ increases, SUs can transmit at higher power levels and $\RLB^*$ increases, until  $\RLB^*$ reaches its maximum value. Increasing $\Ibar$ any further, beyond the knee point in Fig. \ref{C_3Nu_Iav2}, does not increase $\RLB^*$. This is because for large $\Ibar$, transmit power levels are restricted by the amount of harvested and stored energy in the battery (and not by the AIC). Therefore, increasing $\Ibar$ beyond the knee point has no effect on $\RLB^*$. Moreover, for small $\Ibar$ where the AIC is active, increasing $\rho$ has no effect on $\RLB^*$. On the other hand, for large $\Ibar$, when $\rho$ increases, $\RLB^*$ increases.
%
%
%
%
\begin{figure}[!t]
	\centering       			
	\includegraphics[width=60mm]{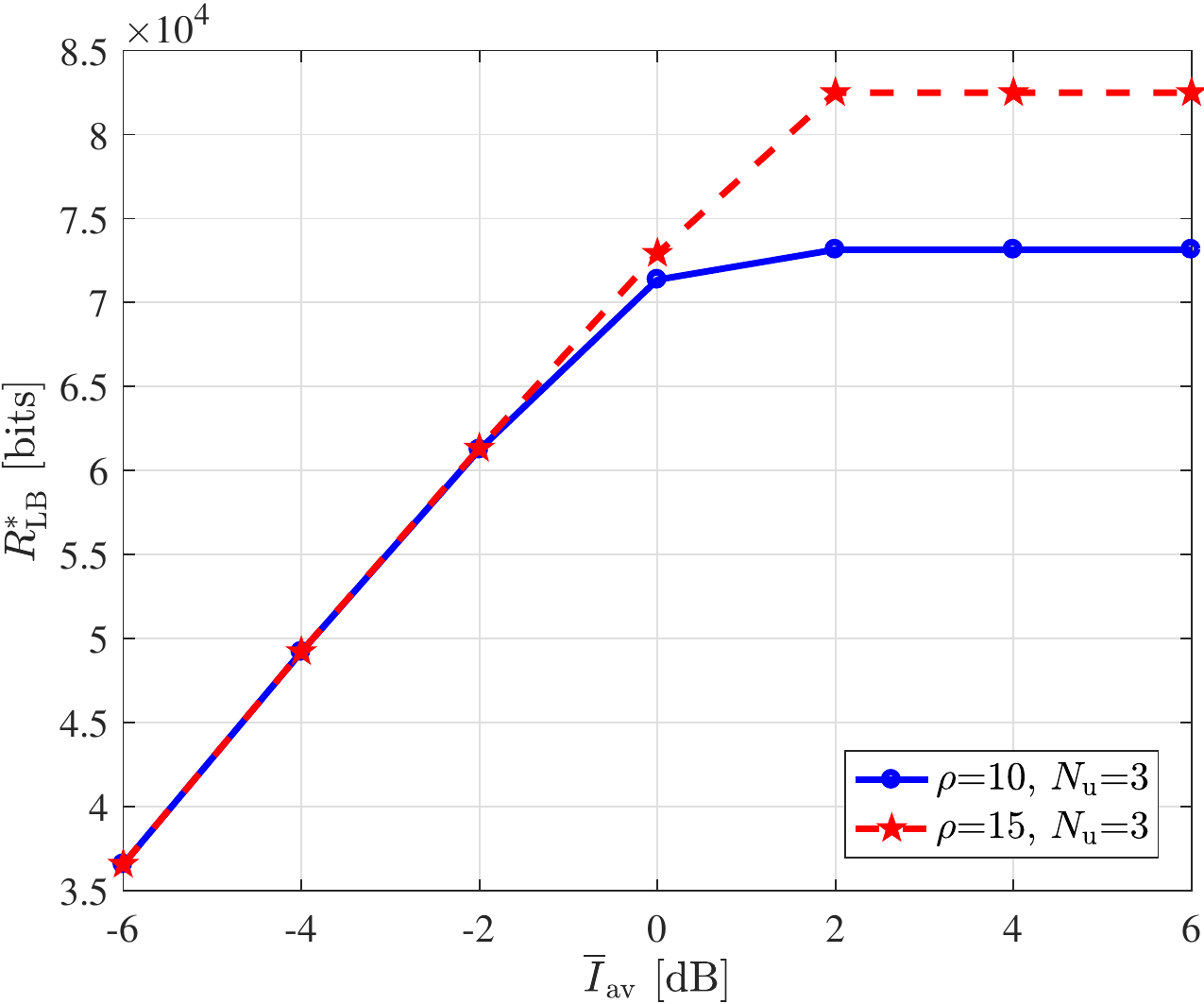}
	\caption{$\RLB^*$ versus $\Ibar$ for $\Nu=3, K=80$.} 
	\label{C_3Nu_Iav2}      
	\vspace{-0mm}
\end{figure}
%
\par Considering SU$_1$, Fig \ref{Pout_vs_K_fig} depicts $P_{b_1}^{\rm{Out}}$ of this user versus $K$ where the optimization variables $\Omega_1$ and $\theta_1$ are obtained by solving \eqref{Prob1} and maximizing $\RLB$ and then substituting the optimized variables in \eqref{Pr_battery_out} to calculate $P_{b_1}^{\rm{Out}}$. We observe that increasing $K$ leads to a lower $P_{b_1}^{\rm{Out}}$.
%
%
%
%
\begin{figure}[!t]
\centering       			
\includegraphics[width=60mm]{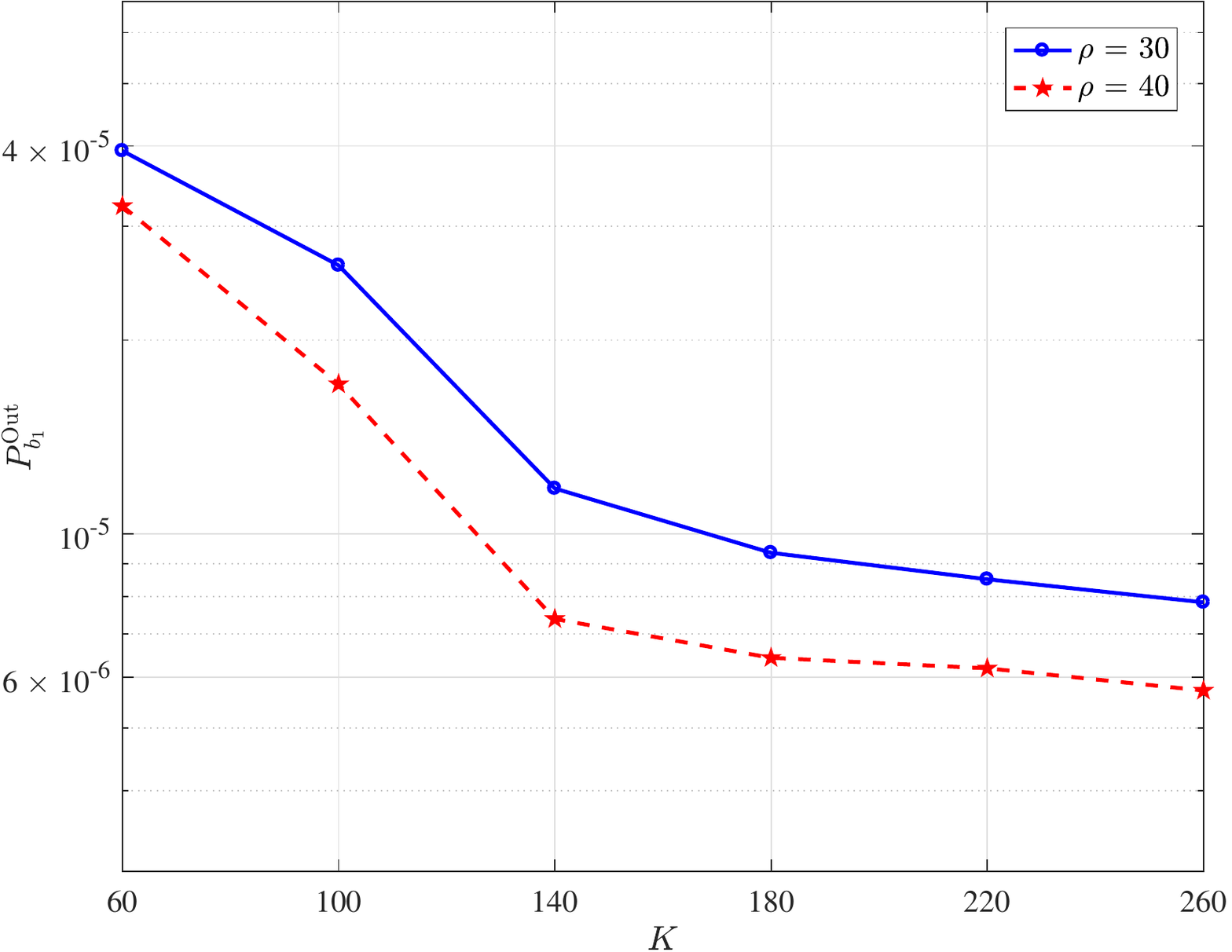}
\caption{$P_{b_1}^{\rm{Out}}$ for SU$_1$ versus $K$  when $\Ibar=2\,$dB.} 
\label{Pout_vs_K_fig}      
\end{figure}
%
\par We define the transmission outage probability $P_{\alpha_n}^{\rm{Out}}$ as the probability of SU$_n$ not being able to transmit data to the AP (due to either a weak SU$_n$--AP link with small fading coefficient or insufficient amount of stored energy at the battery). We have 
%
\begin{align}\label{Pout1}
P_{\alpha_n}^{\rm{Out}} =\Pr(P_{n} = 0 | \Hhaton) = \omega_{0,n} \Pr(P_{n} = 0 | \Hhaton, \Ho) \nonumber\\
+ \omega_{1,n} \Pr(P_{n} = 0 | \Hhaton, \Hl),
\end{align}
%
where
%
\begin{align} \label{Pout2}
\Pr(P_{n}& = 0 | \Hhaton, \Heps) \nonumber \\
= &\sum_{k=0}^{\alphatra} \zeta_{k,n} \Pr(\alpha_{k,n} = 0 | \Hhaton, \Heps, {\cal B}_n \leq \alphatra) \nonumber\\
+ &\sum_{k=\alphatra \! + \! 1}^{K} \zeta_{k,n} \Pr(\alpha_{k,n} = 0 | \Hhaton, \Heps, {\cal B}_n \! \geq \! \alphatra \! + \! 1). 
\end{align}
%
Substituting  \eqref{Prob_alpha} and \eqref{Pout2} in \eqref{Pout1} we get 
%
\begin{equation}\label{Pout3}
P_{\alpha_n}^{\rm{Out}}= \sum_{k=0}^{\alphatra} \zeta_{k,n} + \sum_{k=\alphatra+1}^{K} \zeta_{k,n} Y_{k,n}.
\end{equation}
%
%
%
\noindent Fig. \ref{Pout_Iav_fig}  shows $P_{\alpha_1}^{\rm{Out}}$ for SU$_1$ versus $\Ibar$  where the optimization variables  $\Omega_1$ and $\theta_1$ are obtained by solving \eqref{Prob1} and maximizing $\RLB$ and then substituting the optimized variables in \eqref{Pout3} to compute $P_{\alpha_1}^{\rm{Out}}$. Starting from small $\Ibar$, as $\Ibar$ increases, SUs can transmit at higher power levels and $P_{\alpha_1}^{\rm{Out}}$ decreases, until  $P_{\alpha_1}^{\rm{Out}}$ reaches its minimum value. Increasing $\Ibar$ any further, beyond the knee point in Fig. \ref{Pout_Iav_fig}, does not reduce $P_{\alpha_1}^{\rm{Out}}$. This is because for large $\Ibar$ transmit power levels are restricted by the amount of harvested and stored energy in the battery (and not by the AIC). Therefore, increasing $\Ibar$ beyond the knee point has no effect on $P_{\alpha_1}^{\rm{Out}}$. 
%
%
%
\begin{figure}[!t]
\centering       			
\includegraphics[width=60mm]{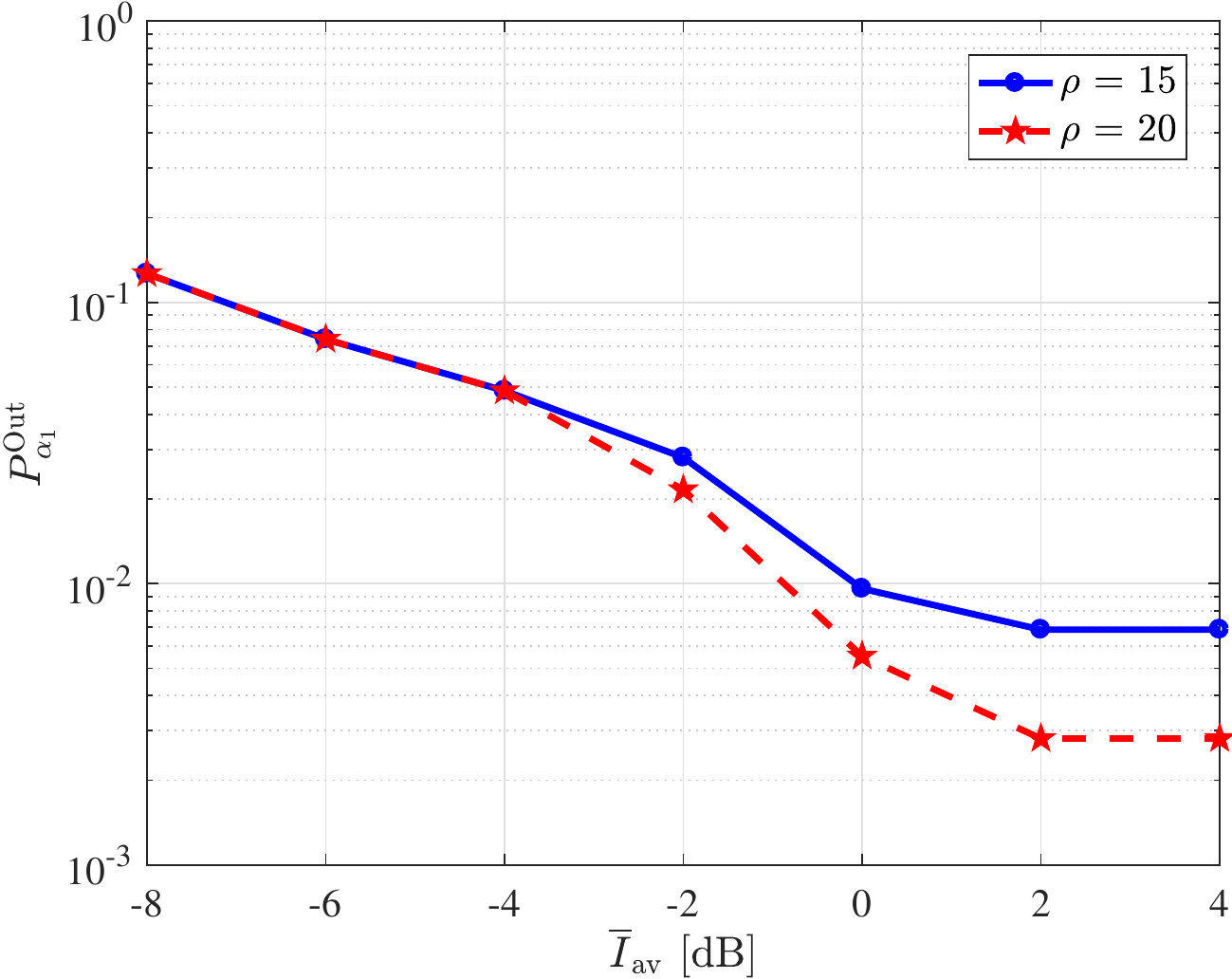}
\caption{$P_{\alpha_1}^{\rm{Out}}$ for SU$_1$ versus $\Ibar$ for SU$_1$ when $K=100$.} 
\label{Pout_Iav_fig}      
\vspace{-0mm}
\end{figure}
%
%
\section{Conclusion} \label{Conclu}
We considered an opportunistic  CR network,  consisting of $\Nu$  SUs and the AP, that can access a spectrum band licensed to a primary network. Each SU is capable of harvesting energy from ambient energy sources, and is equipped with a finite size  battery, for storing its harvested energy. The SUs operate under a time-slotted scheme, where each time slot consists of: spectrum sensing phase, channel probing phase, and data transmission phase.  
To achieve a balance between the energy harvesting and the energy consumption, we proposed a parametrized power control strategy that allows each SU to adapt its power, according to the received feedback information from the AP regarding its link fading coefficient and its stored energy in the battery. Modeling  the randomly arriving energy packets during a time slot as a Poisson process, and the dynamics of the battery as a finite state Markov chain, we established a lower bound on the achievable sum-rate of SUs--AP links, in the presence of both spectrum sensing and channel estimation errors. We optimized the parameters of the proposed power control strategy, such that the derived sum-rate lower bound is maximized, subject to the AIC. We validated our analysis via Matlab simulations and explored spectrum sensing-channel probing-data transmission trade-offs. We also illustrated how the AIC, the harvesting parameter, and the battery size impact the sum-rate, as well as transmission outage probability.
%
%
%
%
%
%
\bibliographystyle{IEEEtran}
\bibliography{EH_Ref}
%
\end{document}